\documentclass[twocolumn,showpacs,amsmath,amssymb,prb]{revtex4}

\usepackage{graphicx}	
\usepackage{dcolumn}	
\usepackage{bm}		
\usepackage[T1]{fontenc}

\newcommand{\YCBCO}{\mbox{Y$_{0.5}$Ca$_{0.5}$BaCo$_{4}$O$_{7}$}}
\newcommand{\YBCO}{\mbox{YBaCo$_{4}$O$_{7}$}}
\newcommand{\grad}{\ensuremath{^\circ}}

\begin{document}

\title{Spin dynamics, short range order and spin freezing in Y$\bm{_{0.5}}$Ca$\bm{_{0.5}}$BaCo$\bm{_{4}}$O$\bm{_{7}}$}

\author{J. R. Stewart}
\affiliation{ISIS Facility, Rutherford Appleton Laboratory, Didcot, OX11 0QX, UK}
\email{ross.stewart@stfc.ac.uk}
\author{G. Ehlers}
\affiliation{Neutron Scattering Science Division, Oak Ridge National Laboratory, Bldg. 8600, Oak Ridge, TN 37831-6475, USA}
\author{H. Mutka}
\author{P. Fouquet}
\affiliation{Institut Laue-Langevin, 6 rue Jules Horowitz, BP 156, 38042 Grenoble Cedex 9, France}
\author{C. Payen}
\affiliation{Institut des Mat{\'e}riaux Jean Rouxel (IMN), Universit{\'e} de Nantes-CNRS, BP 32229, 44322 Nantes Cedex 3, France}
\author{R. Lortz}
\affiliation{Department of Condensed Matter Physics, University of Geneva, 24 Quai Ernest-Ansermet, CH-1211 Geneva 4, Switzerland}
\altaffiliation{present address: Department of Physics, The Hong Kong University of Science and Technology, Clear Water Bay, Kowloon, Hong Kong }

\date{\today}


\begin{abstract}
\YCBCO\ was recently introduced as a possible candidate for capturing some of the predicted classical spin kagome  ground state features.   Stimulated by this conjecture we have taken up a more complete study of the spin correlations in this compound with neutron scattering methods on a powder sample characterized with high--resolution neutron diffraction and the temperature dependence of magnetic susceptibility and specific heat.   We have found that the frustrated near neighbor magnetic correlations involve not only the kagome planes but concern the full Co sub--lattice as evidenced by the analysis of the wavevector--dependence of the short range order.  We conclude from our results that the magnetic moments are located on the Co sub--lattice as a whole and that correlations extend beyond the two--dimensional kagome planes.  We identify intriguing dynamical properties, observing high--frequency fluctuations with a lorentzian line--width $\Gamma\le{20}$~meV at ambient temperature.   On cooling a low--frequency ($\sim1 $~meV) dynamical component develops alongside the high--frequency fluctuations, which eventually becomes static at temperatures below $T\approx{50}$~K.   The high--frequency response with an overall line--width of  $\sim$ 10~meV  prevails at $T\le{2}$~K coincident with a fully elastic short--range ordered contribution.   
\end{abstract}

\maketitle


\section{Introduction}
\label{Introduction} 

An ensemble of magnetic moments (``spins'') residing on a two--dimensional \textit{kagome} lattice has been known for a long time as a prototype for topological magnetic frustration.~\cite{syozi51,ram01}  In recent years, a number of materials has been synthesized where \textit{S}=1/2 or \textit{S}=3/2 spins reside on such a lattice or have a sub--lattice containing kagome layers.  If the couplings are  sufficiently two--dimensional in character, long--range order at low temperature is suppressed and the effects of frustration can be studied experimentally.  Topical materials in this class include \textit{S}=1/2 systems such as volborthite, Cu$_3$V$_2$O$_7$(OH)$_{2}\cdot{2}$H$_2$O,~\cite{hiroi01,fuka03,bert05} and herbertsmithite, ZnCu$_3$(OH)$_6$Cl$_2$,~\cite{helt07,rigol07,imai08,olar08,DeVries09} as well others with higher spin quantum number, such as the \textit{S}=5/2 jarosites AFe$_{3}$(OH)$_{6}$(SO$_{4}$)$_{2}$,~\cite{wills00, matan06} and the double kagome layer (termed \textit{pyrochlore slab}) compounds with \textit{S}=3/2 -- SrCr$_{9x}$Ga$_{12-9x}$O$_{19}$ (SCGO)~\cite{obradors88,ram90,broh90,lee96a,mut06} and Ba$_2$Sn$_2$ZnCr$_{7x}$Ga$_{10-7x}$O$_{22}$ (BSZCGO).~\cite{hag01,bonnet04, bono04a, bono04c}  Due to complicated chemical composition and crystal structure the low temperature properties of these systems are complex and sometimes controversial.  Deviations from the simplest nearest neighbor isotropic interaction picture, slight deviations from the \textit{ideal} frustrated topology of the magnetic sub--lattice and the role of defects and impurities can influence the properties in a way that may be difficult to clarify.  A long term goal in the field has been to synthesize compounds that would exhibit and therefore confirm  the expected quantum (Ref.~\onlinecite{sindz09} and references therein) or classical~\cite{huse92,reimers93,robert08,zhito08,henley09} properties revealed in theoretical and numerical studies of the nearest neighbor Heisenberg hamiltonian. 

Recent studies of ``114'' cobaltite compounds with the composition YBaCo$_4$O$_x$ $(7\lesssim{x}\lesssim{8})$ and substituted variants have revealed the interesting frustrated properties and intricate chemistry of these compounds.~\cite{valldor02,valldor04a,valldor04b,karppi05,soda06,maignan06,chapon06,chmai08,tsipis09,holl09,manuel09}  The substitutionally disordered system, \YCBCO,~\cite{valldor06} was proposed as a new kagome material involving Co$^{2+}$ (\textit{S}=3/2) ions.~\cite{schweika07}  A combination of different arguments involving charge balance, local Co ion site symmetry and possible oxygen deficiency was used to propound \YCBCO\ as a two--dimensional spin system in which high--spin (\textit{S}=3/2) Co$^{2+}$ ions are arranged in kagome planes (labelled Co2 on the $6c$ Wyckoff position), with additional low--spin (\textit{S}=0) Co$^{3+}$ ions preferentially residing in intermediate layers (labelled Co1 on the triangular planar $2a$ Wyckoff position), separating the kagome planes.  Quantitative modeling of the reported cold neutron polarized neutron diffraction data led Schweika and co--workers to the conclusion that the observed features match the classical kagome ordered state expectations~\cite{schweika07} allowing for the ``weathervane'' defects typical of the low--energy response of the system.~\cite{matan06}  Motivated by this observation we have undertaken a more complete neutron scattering examination aiming at better understanding of the static and dynamic magnetic response, in the context of theoretical work on the dynamics and ordering of the classical kagome system.~\cite{robert08,zhito08,henley09}

It was reported earlier that \YCBCO\ is characterized by a large magnitude Curie--Weiss constant, $\Theta_{\rm{CW}}\sim{-2200}$~K, and the apparent absence of a magnetic ordering transition down to $T\sim{1.2}$~K,~\cite{valldor06,schweika07} suggesting a significant degree of frustration.  Accordingly the magnetic properties of \YCBCO\ appear very different from the parent compound \YBCO\ which shows a transition to a state with long range magnetic order below $T=110$~K, at a reasonably high temperature, but nevertheless below the mean field expectation based on the value $\Theta_{\rm{CW}}\sim{-510}$~K found for this compound.~\cite{chapon06}One might argue that the difference with respect to the non--substituted case is associated with the important effect of heterovalent substitution of Ca$^{2+}$ for Y$^{3+}$ and the resulting enhancement of geometric frustration due to decrease of dimensionality.  The magnitude of the Curie--Weiss constant indicates the expected energy scale of the dynamic response, and accordingly we have chosen to investigate the dynamic response of \YCBCO\ over a broader energy range beyond the cold neutron regime investigated earlier.~\cite{schweika07}  

The  presentation of our results is organized as follows.  After a short account of the experimental techniques in section~\ref{Experimental} we present an overview of the characteristics of the sample examined in section~\ref{bulk}.  We present a structural study of \YCBCO\ using high--resolution neutron powder diffraction in section~\ref{hrpd}, then in the section~\ref{polar} we present new polarized neutron diffraction data, found to be in quantitative agreement with those reported earlier.~\cite{schweika07}  However, our analysis of the spatial spin correlations suggests that all Co ions carry moments and participate in the magnetic response.  Furthermore, in section~\ref{inel} we present a neutron spectroscopic investigation which indicates a dynamical magnetic response reaching over a wide range of time and energy scales, and also revealing a freezing of the system when it is cooled to low temperatures $T~\le$~50~K, in agreement with our bulk magnetic observations. We show that due to the broad extent of the dynamics, polarized neutron diffraction using cold neutrons misses a significant portion of the fluctuating moment, reinforcing the conclusion that all Co atoms are magnetic and that  \YCBCO\ is not a classical kagome system.


\section{Experimental methods}
\label{Experimental}

A powder sample of \YCBCO\ (mass $\sim{8}$~g) was prepared in a manner similar to that used previously.~\cite{valldor06}  
High--purity Y$_2$O$_3$, CaCO$_3$, BaCO$_3$, and Co$_3$O$_4$ were mixed in stoichiometric ratios and heated in air at 1100~{C} with several intermediate grindings.  After the final firing at 1100~{C}, the sample was quenched to room temperature to minimize the possibility of excess oxygen stoichiometry in the sample. This quenching from high temperatures should also lead to a random substitution of Ca for Y. X--ray powder diffraction was performed with Cu--$K_{\alpha}$ radiation, showing that the sample was well crystallized and single--phase within the sensitivity of the technique. The refined cell parameters at room temperature were consistent with values reported earlier by Valldor,~\cite{valldor06} although we shall see below that their reported structure (hexagonal $P6_{3}mc$ space group) does not match our observations.

Measurements of DC and AC susceptibility  were performed in a Quantum Design MPMS between $T=2$~K and 400~K.
The specific heat data were taken in two sets, one over a range 4~K~$\le{T}\le$~100~K without applied field, another over a range 20~K~$\le{T}\le$~300~K with applied field up to $\mu_{0}H=14$~T.  The lower T data were obtained using a semi-adiabatic method (MagLabVSM, Oxford Instruments) and were corrected for the sample holder and grease contributions.  The high temperature specific heat from 20~K to 300~K was measured with an high precision continuous heating adiabatic calorimeter.

The neutron scattering measurements were made at the Institut Laue--Langevin (ILL) in Grenoble, France, and the ISIS Pulsed Neutron facility in Oxfordshire, UK.  A high--resolution neutron powder diffraction structural study was performed at the HRPD diffractometer at ISIS.  An \textit{xyz} neutron polarization analysis diffraction study was performed using the D7 diffuse scattering instrument at the ILL, with an incident wavelength of 4.8~{\AA}. Time--of--flight (TOF) inelastic neutron spectra were measured using the thermal spectrometer IN4 at the ILL, with various settings of the incident wavelength ($\lambda_{i}=1.1$~{\AA}, 1.3~{\AA}, 2.2~{\AA}) and instrumental resolution in the range between $\Delta(\hbar\omega)\approx{3.5}$~meV to 0.7~meV (FWHM).  Additional inelastic TOF data with higher resolution $\Delta(\hbar\omega)\approx$~100~$\mu$eV (FWHM) were taken using the IN5 cold neutron spectrometer at the ILL operating at $\lambda_{i}=5.0$~{\AA}.  Elastic scans with 1 $\mu$eV resolution were recorded on the backscattering instrument IN16 at the ILL.  In these experiments, data were corrected for neutron detection efficiency and background scattering using standard procedures.  For further investigation of the slow dynamics a neutron spin echo (NSE) experiment was performed using the multidetector spin echo spectrometer IN11C at the ILL, with a neutron wavelength of 5.5~{\AA} (bandwidth 15\%), giving a dynamic range from $5\times{10}^{-12}$~s to $2\times{10}^{-9}$~s and simultaneous coverage of a $Q$ range from 1~{\AA}$^{-1}$ to 1.6~{\AA}$^{-1}$.  A sample of Ho$_{0.7}$Y$_{1.3}$Ti$_2$O$_7$ measured at 1.5~K was used to measure the instrumental resolution in this experiment.~\cite{ehlers06}

D7 and IN11C use cold neutrons with low incident energy ($E_{i}\approx{3.5}$~meV in the experiments reported here) and the supermirror analyzer devices used on these spectrometers effectively limit the energy transfer acceptance for neutron energy gain.~\cite{Stewart00}  This means that the effective \textit{energy transfer window} of these measurements covers the range --10 meV $\lesssim\hbar\omega\lesssim{3}$~meV.  
 
\begin{figure}
	\includegraphics[width=3.0in]{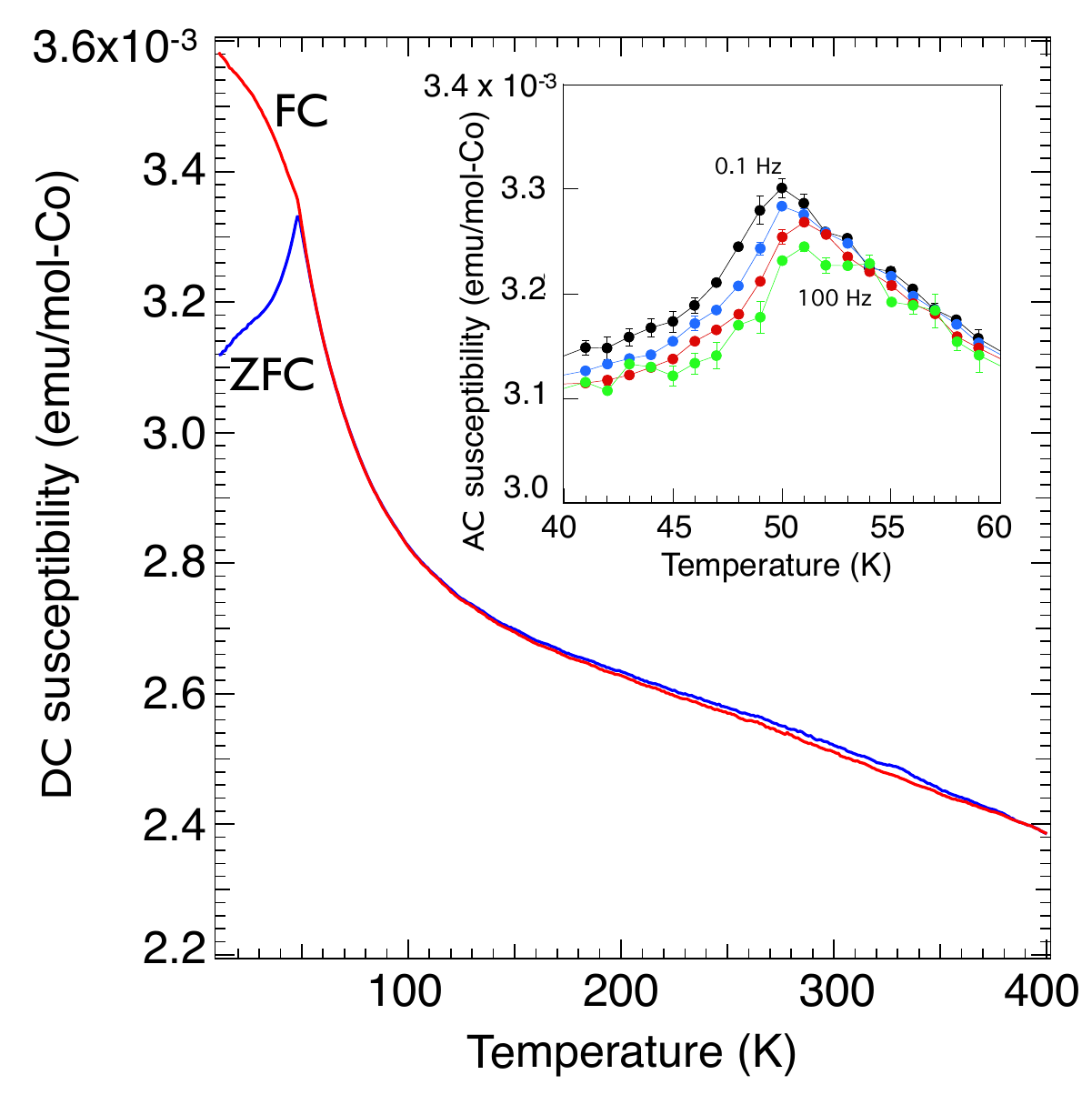}
	\caption{(Color online) The temperature dependent DC magnetic susceptibility at $\mu_{0}H=100$~G (main figure).  Zero--field--cooled (ZFC) and field--cooled (FC) branches are shown  The inset shows the shift to higher temperature of the cusp in the AC susceptibility as a function of frequency, measured using a primary AC field of $\mu_{0}H=3$~G (rms) for $f$ = 0.1 (black), 1 (blue), 10 (red) and 100 Hz (green).}
	\label{SuscGraph}
\end{figure}


\section{Experimental Results}
\label{Results}


\subsection{Susceptibility and specific heat}
\label{bulk}

\begin{figure}
	\includegraphics[width=3.0in]{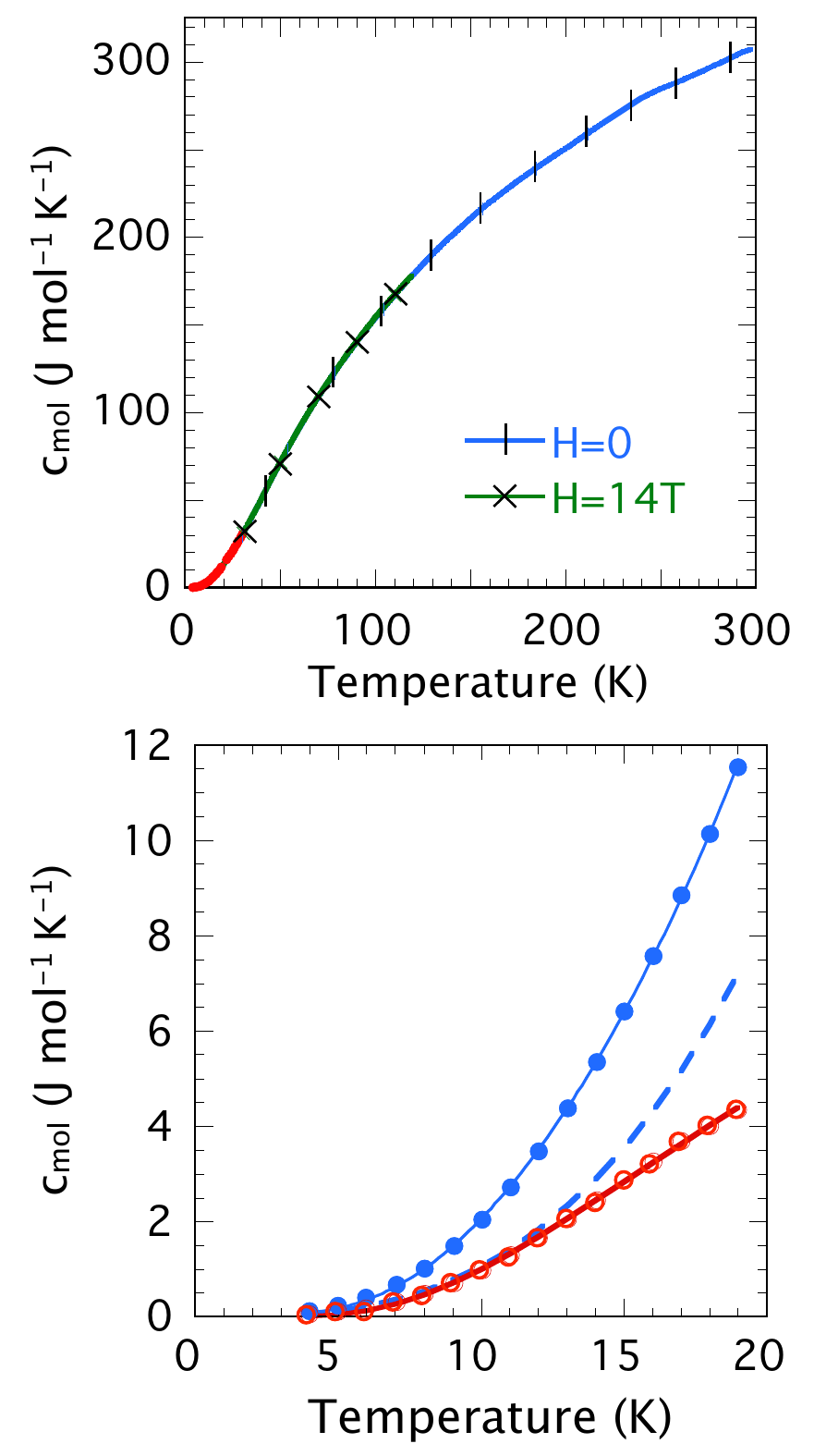}
	\caption{(Color online) The temperature dependence of the specific heat. (top panel)  Data measured in zero magnetic field of  between 20~K $<{T}<$ 300~K (blue, vertical lines), and in a magnetic field of $\mu_{0}H=14$~T between 20~K $<{T}<$ 120~K (green, crosses). Note that these two are totally overlapping indicating no field dependence over the $T$--range concerned. The red points show the low--$T$ portion in zero field. \\ (lower panel) The low--$T$ data in zero field (blue solid circles) fitted to a Debye phonon--term and an exponentially activated contribution $\propto\exp(-E_{a}/k_{B}T)$ (blue solid line) as well as the extracted purely Debye part (blue dashed line) and the difference between the data and the Debye contribution (red hollow circles), with the exponential part of the fit (red solid line).}
	\label{cgraph}
\end{figure}

Fig.~\ref{SuscGraph} shows the temperature dependent DC magnetic susceptibility obtained at $\mu_{0}H=100$~G. The main feature is a sharp cusp in the zero field cooled (ZFC) susceptibility at a temperature of $T_g=48$~K concomitant with a bifurcation of the ZFC and field cooled (FC) susceptibilities, indicative of spin freezing.~\cite{binder86}  At the same temperature the AC susceptibility shows a characteristic, weakly frequency dependent cusp as often observed in systems of freezing spins.  There are similarities with the data reported earlier by Valldor,~\cite{valldor06} but our susceptibility data has fewer peaks and is much less structured and smoother (there were at least three clear peaks in Valldor's data set). The third anomaly reported by Valldor at $T=14$~K is missing, indicating that it may depend on the preparation of the specific \YCBCO\ sample.  There is no evidence of Curie--Weiss like behavior up to the maximum temperature measured.

Fig.~\ref{cgraph} shows the measurement of the specific heat at zero field and an applied field of $\mu_{0}H=14$~T.  No visible anomaly appears in the vicinity of $T_{g}$ with or without an applied field, suggesting the absence of a true collective thermodynamic transition.  Hence, no long range order or any low--energy magnon--like excitations associated with such a state can be expected at low temperature $T\le{T}_{g}$.  The trend as $T\to{300}$~K suggests that the specific heat overshoots the high--temperature Dulong-Petit limit (3 $k_{\rm{B}}$ per ion) of the lattice specific heat which coincides with the top of the vertical scale of the plot (324~J mol$^{-1}$K$^{-1}$).  The specific heat at low temperature,  $T\le{20}$~K, is best fitted using two terms; a main contribution that follows a Debye $T^{3}$ dependence and an activated term $\propto\exp(-E_{a}/k_{B}T)$.  Fitting the low--temperature data to this model we find a Debye temperature $\Theta_{\rm{D}}\sim{290}$~K and an activation energy of $E_{a}/k_{B}$=25...30~K, depending on the fitting range.  As a hypothesis we can consider that the exponential increase of the specific heat at low temperature and the overshoot of the Dulong--Petit limit at high temperature are due to a magnetic contribution to the specific heat.  We shall return to this issue below with regard to the the dynamic response obtained from our neutron spectroscopic data. 


\subsection{High-resolution neutron powder diffraction study}
\label{hrpd}

\begin{figure}
	\includegraphics[width=3.3in]{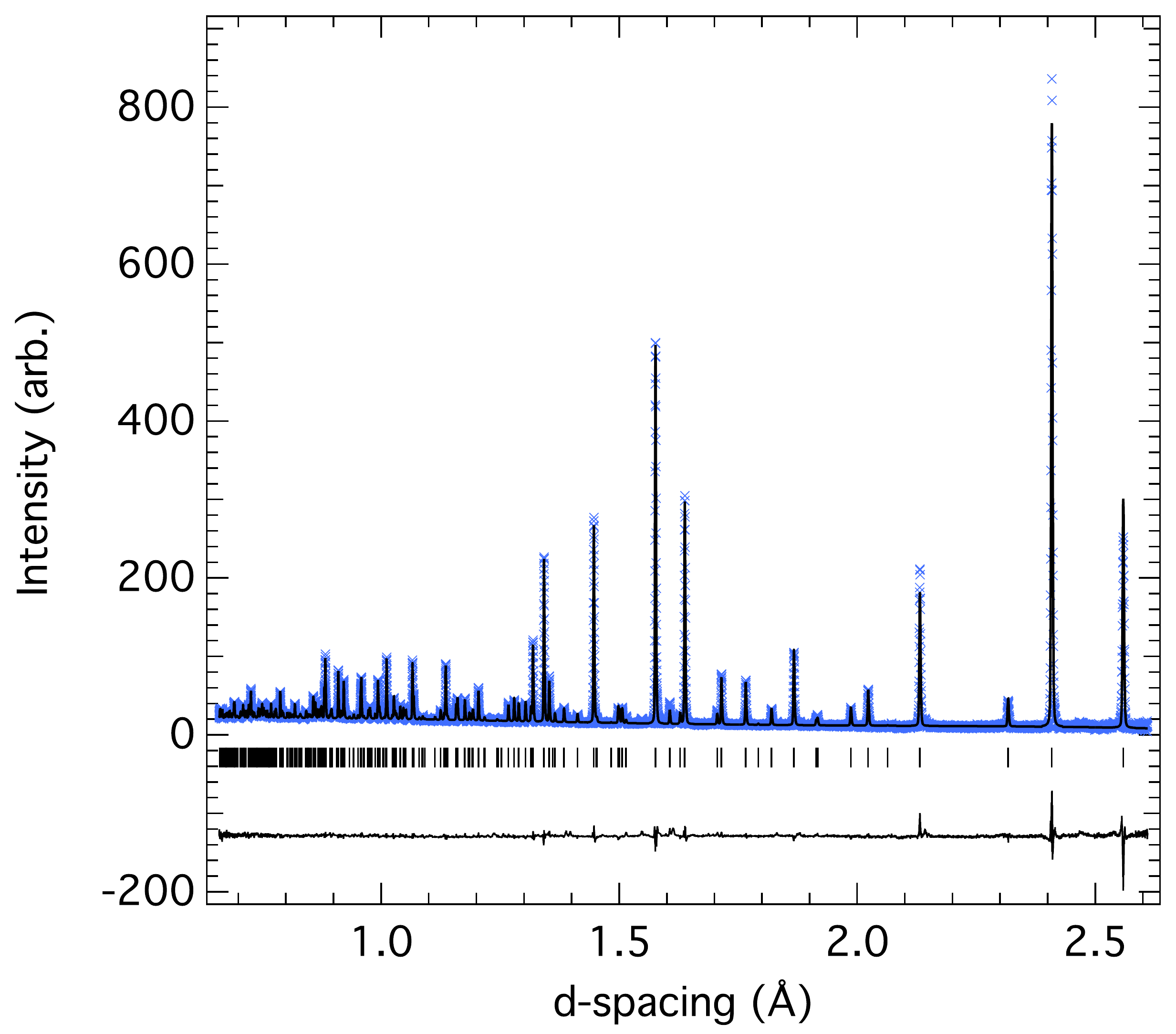}
	\caption{(Color online) Neutron powder diffraction pattern of \YCBCO\ at room temperature, measured using the HRPD diffractometer at ISIS.  A Rietveld refinement model of the data, calculated using {\footnotesize FULLPROF}~\cite{Carvajal93} is shown as a solid line through the data. The difference between the data and model is shown as a solid line below.}
	\label{hrpddata}
\end{figure}

A high--resolution neutron powder diffraction pattern of \YCBCO\ is shown in Fig.~\ref{hrpddata}. Rietveld analysis of the data was carried out using the {\footnotesize FULLPROF} program.~\cite{Carvajal93}  The best model fit to the data was found for the \textit{trigonal} $P31c$ space group, in agreement with most other studies of ``114'' series cobaltites at high temperatures,~\cite{chapon06,huq06,maignan06} but in contrast to the $P6_3mc$ space group found by Valldor using Cu--$K_{\alpha}$ x--ray radiation.~\cite{valldor06}  This is likely due to the fact that x--ray diffraction -- which is much less sensitive to oxygen than neutron diffraction -- was used to solve the structure in that case.

\begin{table}
	\caption{\label{tab1} Atomic parameters for \YCBCO\ at room temperature.  The symmetry is trigonal $P31c$, with cell parameters, $a=6.3075(1)$~\AA~and $c=10.2388(1)$~\AA.  The r-factors were found to be $R_w = 6.05$, $R_{\rm{Bragg}}=5.52$ and a goodness of fit (GoF) $= R_w/R_{\rm{exp}} = 1.83$.}  
	\begin{ruledtabular}
		\begin{tabular}{ccccccc}
		Atom  &  $x$ &  $y$ & $z$ & $B_{\rm{iso}}$ (\AA$^2$) & $Occ.$\\
		\hline
		Y		&	2/3			&	1/3			&	0.8744(1)		&	1.11(7)		&	0.554(1) \\
		Ca		&	\textquotedbl 	& 	\textquotedbl	& 	\textquotedbl	&	\textquotedbl	 &	0.446(1) \\
		Ba		&	2/3			&	1/3			&	1/2			&	1.69(6)		&	1 \\
		Co1		&	0			&	0			&	0.4469(1)		&	1.14(5)		&	1 \\
		Co2		&	0.1737(1)		&	0.8389(1)		&	0.6858(1)		&	0.53(1)		&	3 \\
		O1-1	&	0.494(2)		&	0.512(2)		&	0.7630(8)		&	1.29(6) 		&	1.55(3) \\
		O1-2	&	0.509(2)		&	0.486(2)		&	0.7212(9)		&	\textquotedbl	&	1.45(3) \\
		O2		&	0			&	0			&	0.2523(5)		&	0.501(5)	 	&	1 \\
		O3-1	&	0.1340(6)		&	0.8142(7)		&	0.5015(5)		&	0.61(5)		&	1.85(3) \\
		O3-2	&	0.494(2)		&	0.512(2)		&	0.7630(8)		&	\textquotedbl 	&	1.15(3) \\
		\end{tabular}
	\end{ruledtabular}
\end{table}

\begin{figure}
	\includegraphics[width=3.3in]{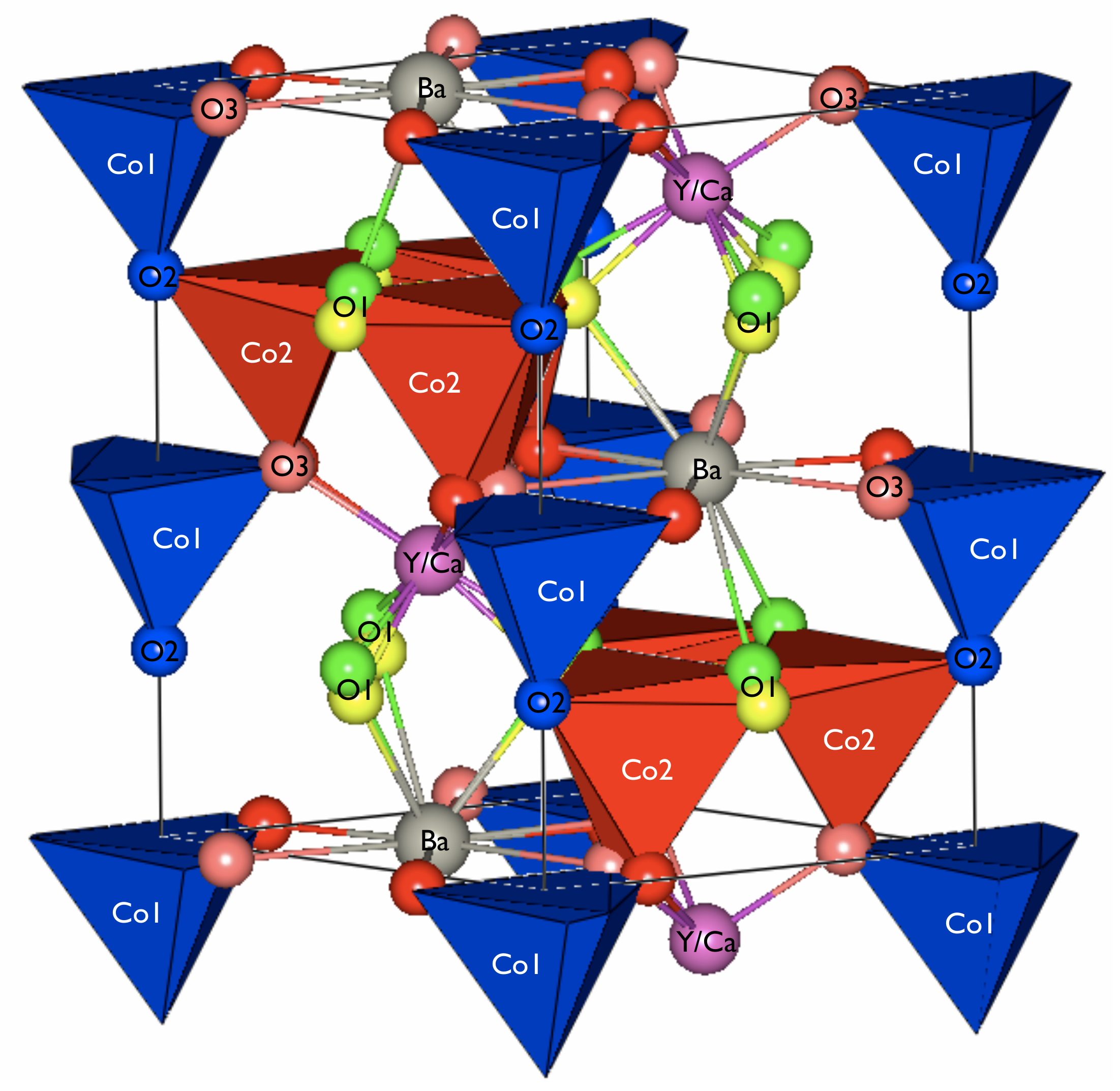}
	\caption{(Color online) Crystal structure of \YCBCO\ at room temperature. Y/Ca atoms in purple, Ba in grey, Co1 centered tetrahedra in blue, Co2 centered tetrahedra in red, O1 in yellow/green, O2 in blue and O3 in red/pink.  The splitting of the O1 and O3 oxygen sites along the Ba-O bonds causes the base of oxygen tetrahedron around the Co2 position (red) to be either parallel to the kagome plane, or to be tilted by an angle of 8.7$^{\circ}$.}
	\label{structure}
\end{figure}

The cell and atomic parameters found from the Rietveld refinement are given in Table~\ref{tab1}, and the crystal structure is depicted in Fig.~\ref{structure}.  Evident from the table is that there are 5 distinct oxygen sites with reasonable thermal parameters in the unit cell, as opposed to 3 in the case of pure \YBCO\, with a splitting of the O1 ($6c$) and O3 ($6c$) sites along the Ba--O bonds. For \YCBCO\ large anomalous thermal parameters were found for oxygen atoms in the $P6_3mc$ model,~\cite{valldor06}  indicating some degree of oxygen disorder.The occupancy of the split O1 and O3 positions was found to be approximately in proportion to the Y--Ca occupation ratio, with the refined Y-Ca occupation found to be 0.55 for Y and 0.45 for Ca.  Thus we surmise that the choice of O1 and O3 position in the unit cell depends on whether that position is coordinated with a Y or Ca ion.  The Y-Ca disorder in the lattice could well have such an effect on the crystal structure, given that; a) the ionic radii of Y$^{3+}$ (1.04~\AA) and Ca$^{2+}$ (1.14~\AA) in an octahedral site are different by $\sim 10$~\%, and; b) in the pure Y--114 sample the CoO$_4$ tetrahedra are aligned to the kagome planes,~\cite{chapon06} while in the pure Ca--114 sample the CoO$_4$ tetrahedra form a corrugated structure out of the kagome plane.~\cite{Caignaert10}  In \YCBCO\ the CoO$_4$ tetrahedron around the kagome Co2 site is either orientated with basal plane aligned with the kagome plane (O11 and O31 positions) - as in the pure Y sample, or oriented with the basal plane at an angle of 8.7$^{\circ}$ to the kagome plane (O12 and O32 positions) - as in the pure Ca sample.  Given that the ratio of site occupation of the O11:O12 and O31:O32 sites is roughly in proportion to the Y:Ca occupation ratio, we conclude that the CoO$_4$ tetrahedra are tilted when the apex is connected to a Ca ion, and aligned when connected to a Y ion. The Co-O distances fall in the range 1.84--1.99~{\AA} and 1.89--2.01~{\AA} for the Co1 site and Co2 site CoO$_4$ tetrahedra respectively.  These distances compare well with those found in \YBCO\ and CaBaCo$_4$O$_7$.~\cite{valldor02,Caignaert10}  Similarly to YBaCo$_4$O$_7$, the CoO$_4$ tetrahedra at the Co2 site are larger than those at the Co1 site, suggesting that the kagome layer of tetrahedra should be preferentially Co$^{2+}$.  Another striking feature of the $P31c$ model concerns the Ba coordination.  The twelve nearest oxygen neighbors are located at distances typical for barium, falling in the range 2.71~{\AA} to 2.99~{\AA}. In other 114 cobaltites, most of the Ba--O distances are more than 3~{\AA} and cannot be classified as true Ba--O bonds.~\cite{valldor02,huq06,Caignaert10}  In these cases, Ba$^{2+}$ is severely underbonded and it has been proposed that some 114 cobaltites undergo a trigonal--orthorhombic structural transition to compensate for this underbonding.~\cite{huq06}  With regard to oxygen stoichiometry, the current data do not allow an accurate estimation.  However, we can say that allowing the oxygen stoichiometry to be a free parameter in the Rietveld refinement did not improve the fit.  Therefore our neutron diffraction data are consistent with an oxygen stoichiometry of 7. 

Concerning the Co2 kagome plane geometry, it is important to note that the geometry is not \textit{perfect} kagome, but rather a lattice of two distinct sizes of corner--sharing equilateral triangle.   Indeed, a perfect kagome lattice is only possible for a general position of type \hbox{$(x, 1-x, y)$} where $x=1/6$.  The side lengths of these two equilateral triangles making up the kagome planes were found to be 3.141~{\AA} and 3.168~{\AA}.    The Co1-Co2 intersite distance is bigger (3.23~{\AA}). The distance between the kagome planes is found to be 5.12~{\AA}.  


\subsection{\textit{xyz} Neutron polarization analysis study}
\label{polar}

\begin{figure}
	\includegraphics[width=3.3in]{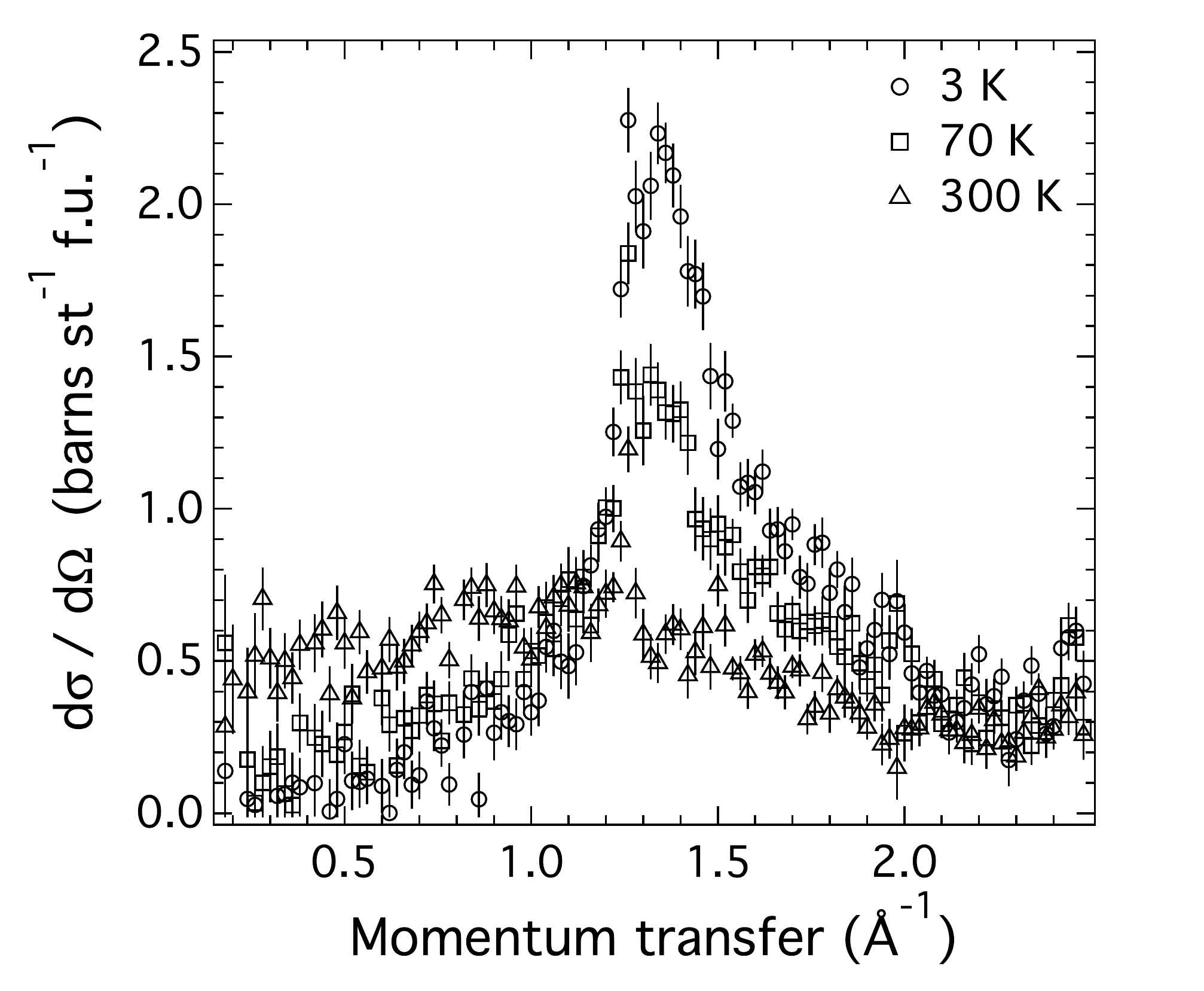}
	\caption{The temperature and wavevector transfer dependence of the \textit{magnetic} scattering (separated from nuclear contributions using \textit{xyz} polarization analysis) of \YCBCO\ between room temperature and 3~K. }
	\label{mag_tempdep}
\end{figure}

Using the technique of \textit{xyz} neutron polarization analysis,~\cite{Stewart09} which allows the separation of the magnetic neutron scattering cross--section from nuclear and nuclear spin contributions, we have measured the magnetic neutron scattering from \YCBCO\ at temperatures between 300 K and 3 K.  Fig.~\ref{mag_tempdep} shows the evolution of the magnetic scattering as a function of temperature and neutron wavevector transfer, $\bm{Q}=\bm{k}_{\rm{i}} - \bm{k}_{\rm{f}}$ where $\bm{k}_{\rm{i}}$ and $\bm{k}_{\rm{f}}$ are the incident and final neutron wavevectors respectively.

\begin{figure}
	\includegraphics[width=3.0in]{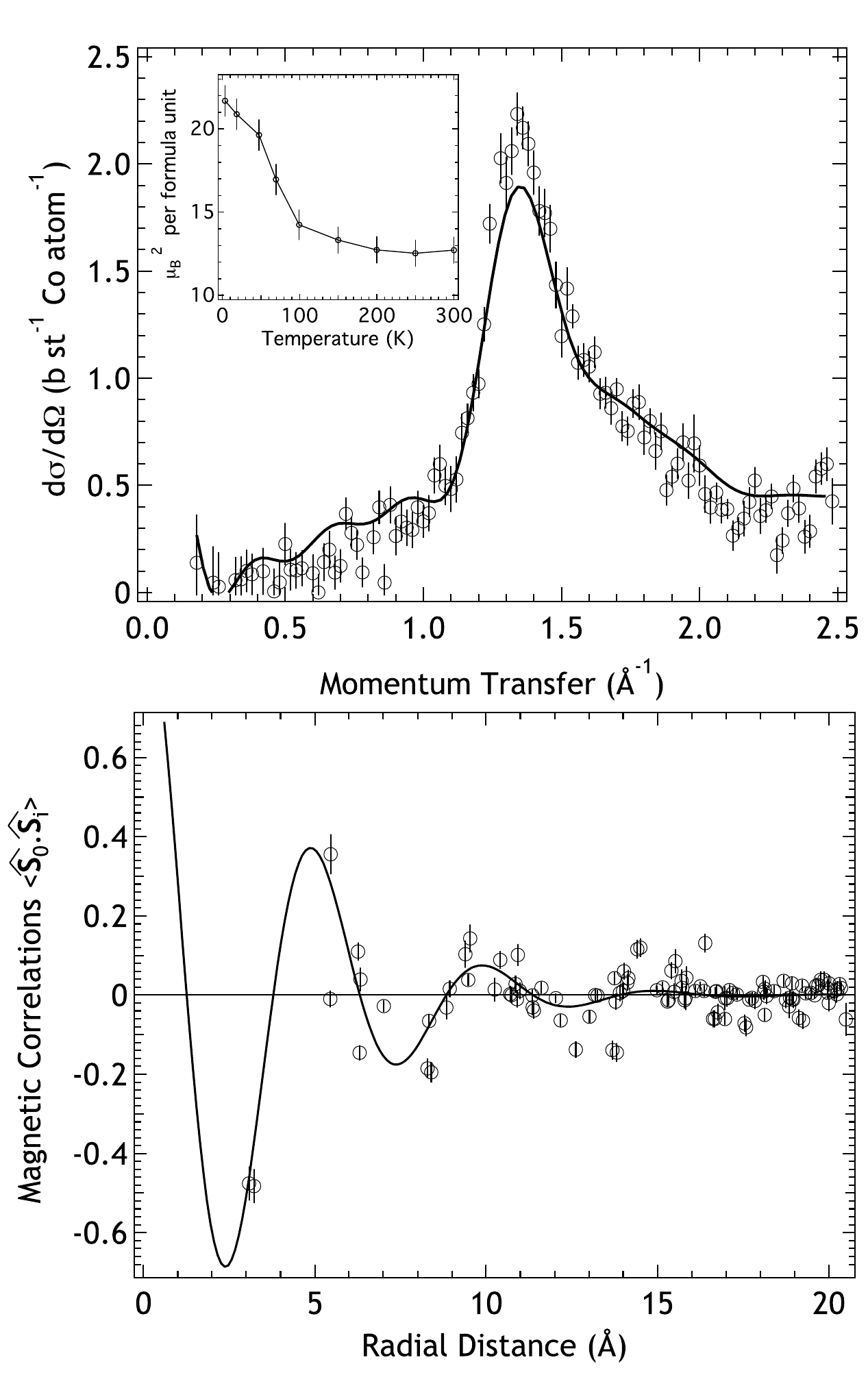}
	\caption{The upper panel shows the experimentally determined scattering cross section $S(Q)$  at $T$=3~K together with an RMC model fit (solid line) assuming non--magnetic Co1 atoms.  The inset shows the temperature dependence of the Co moment per formula unit as determined by our RMC calculations. The lower panel shows the real space, shell to shell spin correlations over distance. The solid line is a guide to the eye.}
	\label{RMC_Correlations}
\end{figure}

The main feature in neutron diffraction is the appearance of a broad peak of diffuse scattering centered around $|\bm{Q}|\sim{1.4}$~{\AA}$^{-1}$, indicating short ranged spin correlations.  These correlations begin to appear at rather high temperature, with evidence of short--range order at the highest temperature measured (300 K).   The magnetic intensity appears to increase monotonically as the temperature decreases.  This indicates that the Co moments are becoming increasingly ``static'' on the timescale of the D7 measurement.  As explained in section~\ref{Experimental} the neutron cross--section measured on D7 is integrated over energy--transfer between the limits --10 meV and +3 meV, implying that the spin--fluctuation rate of the Co moments must be less than $\sim{10}^{12}$ Hz if they are to contribute to the measured scattering cross--section on D7.  The integrated magnetic moment derived from the temperature scan on D7 is shown in the inset of Fig.~\ref{RMC_Correlations}.  Sharp Bragg-peaks may also be seen in the data at $|\bm{Q}|={1.25}$~{\AA}$^{-1}$ and $|\bm{Q}|={2.45}$~{\AA}$^{-1}$.  These peaks, which are independent of temperature, are the first two magnetic Bragg peaks of CoO, which is an antiferromagnet below $T_N=290$~K.  The intensity of the CoO magnetic peaks, in conjunction with the fact that CoO peaks are absent from the powder neutron diffraction data presented in section~\ref{hrpd}, allows us to put an upper limit of 1\% by weight for the CoO impurity level.  

Modeling of the data was undertaken by performing a reverse Monte Carlo (RMC) fit of the D7 data.  Initially, assuming a non-magnetic Co1 $(2a)$ site, as was proposed in the previous study, these moments were excluded from the simulation.  A box of $8\times{4}\times{4}$ unit cells (making a total of 1536 Co2 atoms) was set up with an initially random spin configuration.  In the simulation, the spins were assumed to be fully classical (Heisenberg 3D) and of uniform magnitude.  In each Monte Carlo step, one spin was randomly chosen and rotated by a random angle, and the powder average of the cross section $({d\sigma}/{d\Omega})_{\rm{Mag}}$ was computed using
\begin{eqnarray}
	\label{crossec}
	\left(\frac{d\sigma}{d\Omega}\right)_{\mathrm{Mag}} = 
	\frac{2}{3}\left(\frac{\gamma_nr_0^2}{2}\right)^2f^2\left(Q\right) \mu^2
	\times \nonumber \\
	\left[1+\sum_n\left\langle\widehat{\bm{S}}_0\cdot\widehat{\bm{S}}_n\right\rangle
	Z_n\frac{\sin Qr_n}{Qr_n}\right]
\end{eqnarray}
where the magnetic pre-factor $2/3(\gamma_nr_0^2/2)^2$ = 0.049 barns, $f^2(Q)$ is the Co $(2+)$ squared magnetic form factor, $\mu^2$ is the squared magnetic moment per Co atom, $r_n$ and $Z_n$ are the distances and coordination numbers of the $n^{\rm{th}}$ near neighbor Co coordination shell and $\left\langle\widehat{\bm{S}}_0\cdot\widehat{\bm{S}}_n\right\rangle$ is the average Heisenberg spin--correlation between unit spins on a central Co atom at the origin and the $n^{\rm{th}}$ near neighbor Co coordination shell.  This model was then compared with the data, and a $\chi^2$ goodness of fit parameter calculated. Each RMC step (individual spin rotation) was accepted if it resulted in an improved fit,  and rejected if not. In order to verify that the calculated model spin configuration was independent of the initial random starting position, many individual RMC calculations were performed, and the results averaged.

The scattering cross-section $d \sigma/{d}\Omega$ resulting from the RMC is shown in the upper panel of Fig.~\ref{RMC_Correlations} as a solid line.  The goodness of fit parameter obtained for this model was found to be $\chi^{2}=2.61$. 

\begin{figure}
\includegraphics[width=3.3in]{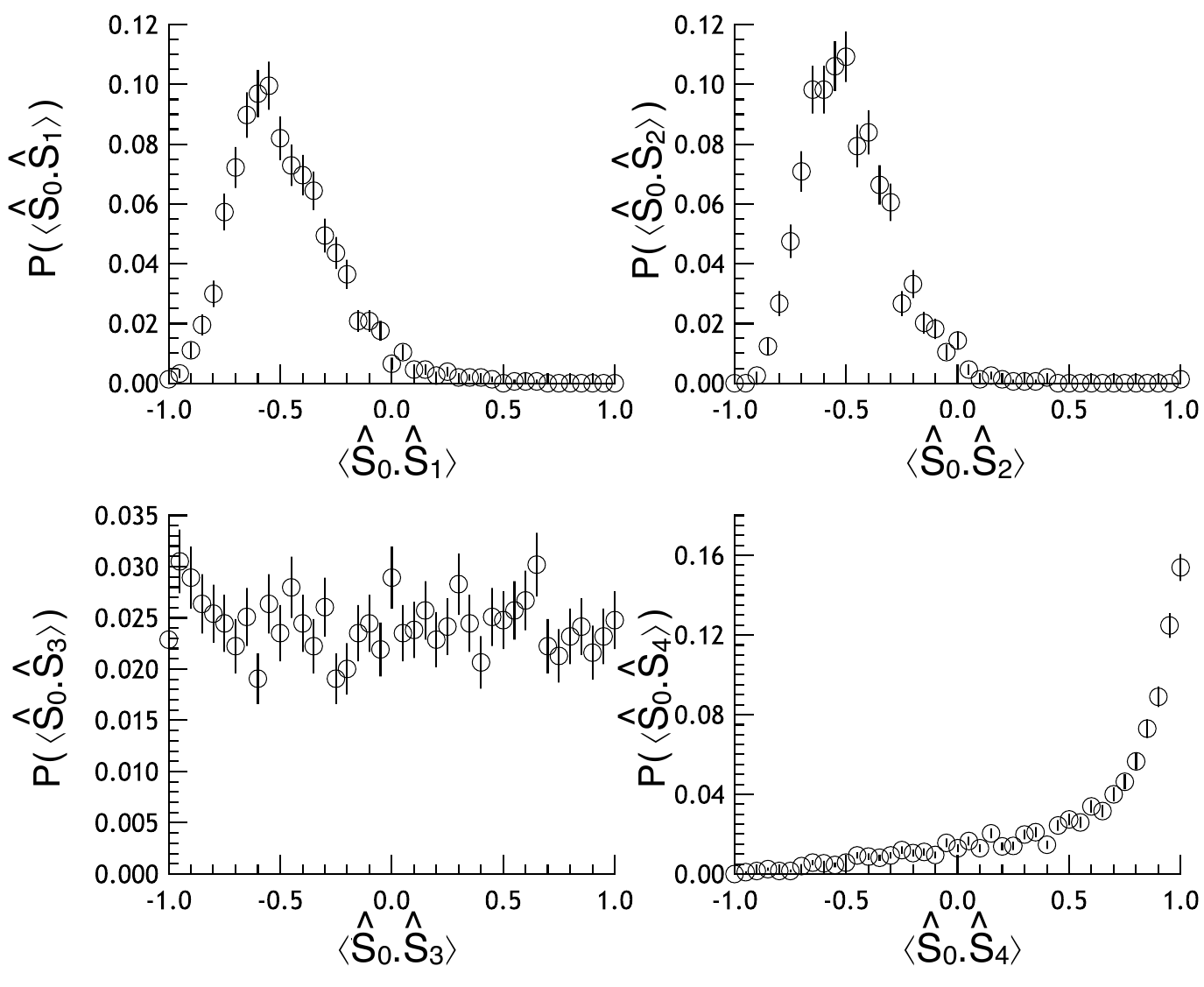}
\caption{The correlation distributions in the first four near neighbor shells from RMC simulation at $T=3$~K, assuming non-magnetic Co1--atoms.  The first two shells are in the kagome plane and show a strong 120\grad\ correlation}
\label{RMC_Shells}
\end{figure}

Fig.~\ref{RMC_Shells} shows the correlation distributions in the first four near neighbor shells resulting from the RMC calculation.  The first two shells (Fig.~\ref{RMC_Shells}, top left) show a peak in the correlation distribution function at $-0.5$, that is, an average angle of 120\grad\ between spins located in the same triangle.  From the oscillation period of the radial correlations (lower panel of Fig.~\ref{RMC_Correlations}) an antiferromagnetic wavevector of 1.22~{\AA}$^{-1}$ is estimated, half way between $Q_0=1.15$~{\AA}$^{-1}$ and $Q_{\sqrt{3}}=1.33$~{\AA}$^{-1}$ (in the notation of Ref.~\onlinecite{schweika07}), which denotes an ordered antiferromagnetic structure with uniform chirality, where the magnetic and atomic unit cells coincide (AF ordering wavevector $\bm{k}_{\rm{AF}}=0$), and a staggered chiral structure, where the magnetic unit cell is expanded by $\sqrt{3}\times\sqrt{3}$ with respect to the atomic unit cell, ($\bm{k}_{\rm{AF}}(1/3,1/3)$).

\begin{figure}
\includegraphics[width=3.0in]{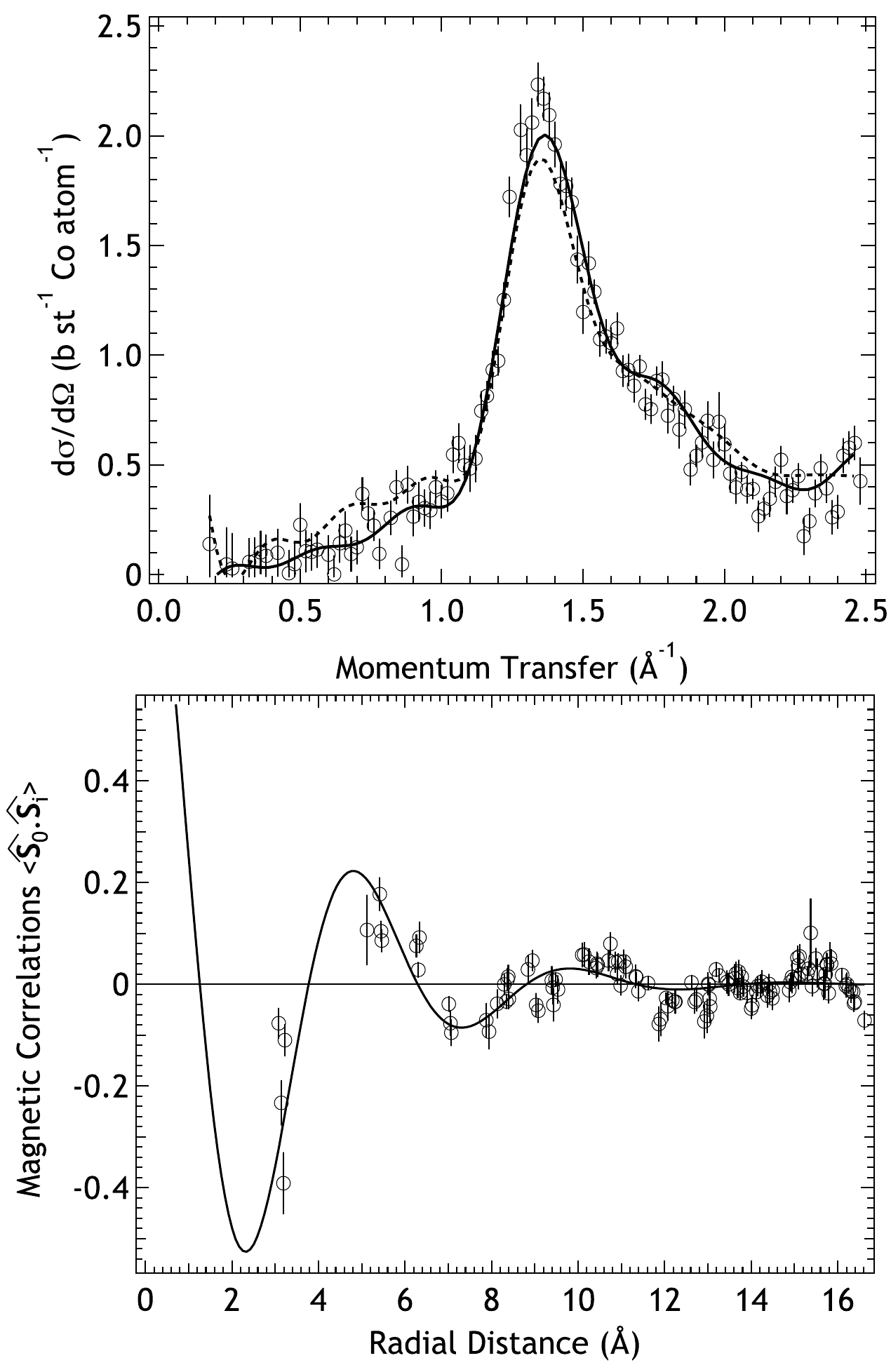}
\caption{The upper panel shows the experimentally determined scattering cross section $S(Q)$  at $T$=3 K together with the RMC simulation assuming both Co1 and Co2 atoms are magnetic (solid line), compared with the previous fit from Fig.~\ref{RMC_Correlations} with the Co1 site excluded (dashed line).  The lower panel shows the real space, shell to shell spin correlations over distance. The solid line is a guide to the eye.}
\label{fullRMC}
\end{figure}

The use of \textit{xyz} polarization analysis enables the scattering cross section to be easily converted to absolute units, allowing a determination of the average magnetic moment, by allowing $\mu^2$ in Eq.~\ref{crossec} to vary.  In this way, we obtain a squared moment of $\mu^{2}={21.6}\pm{0.8}$~$\mu_{B}^{2}$ per formula unit at $T=3$~K.  Note that this moment varies strongly with temperature, as shown in the inset of Fig.~\ref{RMC_Correlations}.  This is a similar number to that found by Schweika and co-workers,~\cite{schweika07} and is the basis of their conjecture that the Co1 $(2a)$ sublattice has zero moment.  Assuming this, then the remaining three Co2 $(6c)$ atoms have a squared-moment of $7.2\pm{0.3}$~$\mu_{\rm{B}}^2$ each.  If we assume that the Co2 atoms are frozen, and that there is no orbital contribution to the moment, then the measured spin per Co would be $S=\mu/g_s=1.34\pm 0.03$.  This is around 90 \% of the expected $S=3/2$ (on the Co2 site) and therefore accounts for the majority of the Co moment.  If, on the other hand, we were to assume that a moment exists on the Co1 sublattice (and that the measured total moment per formula unit were shared out evenly between the Co positions), we would find a squared moment of only $5.4\pm0.2$~$\mu_{\rm{B}}^2$~per Co atom, corresponding to a spin $S=1.16\pm0.02$ or only 64\% of the expected $S=1.81$ (made up of 2.5 \textit{S}=3/2 Co$^{2+}$ and 1.5 \textit{S}=2 Co$^{3+}$ atoms per formula unit~\cite{valldor06}).  This fraction of $\sim{36}\%$ of missing Co moment is, however, readily explainable without the necessity of claiming a low-spin \textit{S}=0 state on the Co1 sublattice.   Any Co moment with a fluctuation rate $\gtrsim{10}^{12}$ Hz will not contribute to the measured scattering cross--section due to the limited energy integration window of the D7 spectrometer, as discussed above.  Therefore, the missing moment fraction could well be indicative of high--frequency magnetic fluctuations, outside the measurement window.  There are many cases of residual high--frequency quantum zero--point fluctuations at low temperatures in frustrated and disordered magnets, and these high--frequency fluctuations often lead to a lower--than--expected integrated magnetic moment, when measured using cold neutrons.~\cite{Cywinski99,DeVries09}  We shall see below that such high frequency fluctuations are indeed observed in \YCBCO.

Given that there is no \emph{a priori} justification for the assumption of a low--spin \textit{S}=0 moment on the Co1 site (see Section~\ref{Discussion}), we now revisit the analysis of the magnetic diffuse scattering, and include these sites in the RMC model.   It is now assumed that both Co sites have the same moment.  The fitted magnetic structure factor, with the Co1 site included, is shown in Fig.~\ref{fullRMC}, together with the previous model with the Co1 site excluded.  
The inclusion of the Co1 site results in a fit of improved quality, with $\chi^{2}=1.69$, indicating that our powder neutron diffraction data seems to indicate the \emph{presence} of a Co1 moment.  We conclude therefore that the observed magnetic scattering cross--section is not consistent with a planar kagome--like magnetic ground state, but rather a 3d antiferromagnetically correlated structure, with magnetic moments on all the Co sites.  This is consistent with the observation of magnetic moments on all of the Co sites in ordered stoichiometric \YBCO,~\cite{chapon06,manuel09} and disordered off--stoichiometric YBaCo$_4$O$_{7.1}$.~\cite{chapon_privcomm}


\subsection{Dynamic magnetic response}
\label{inel}

\begin{figure}
	\includegraphics[width=3.3in]{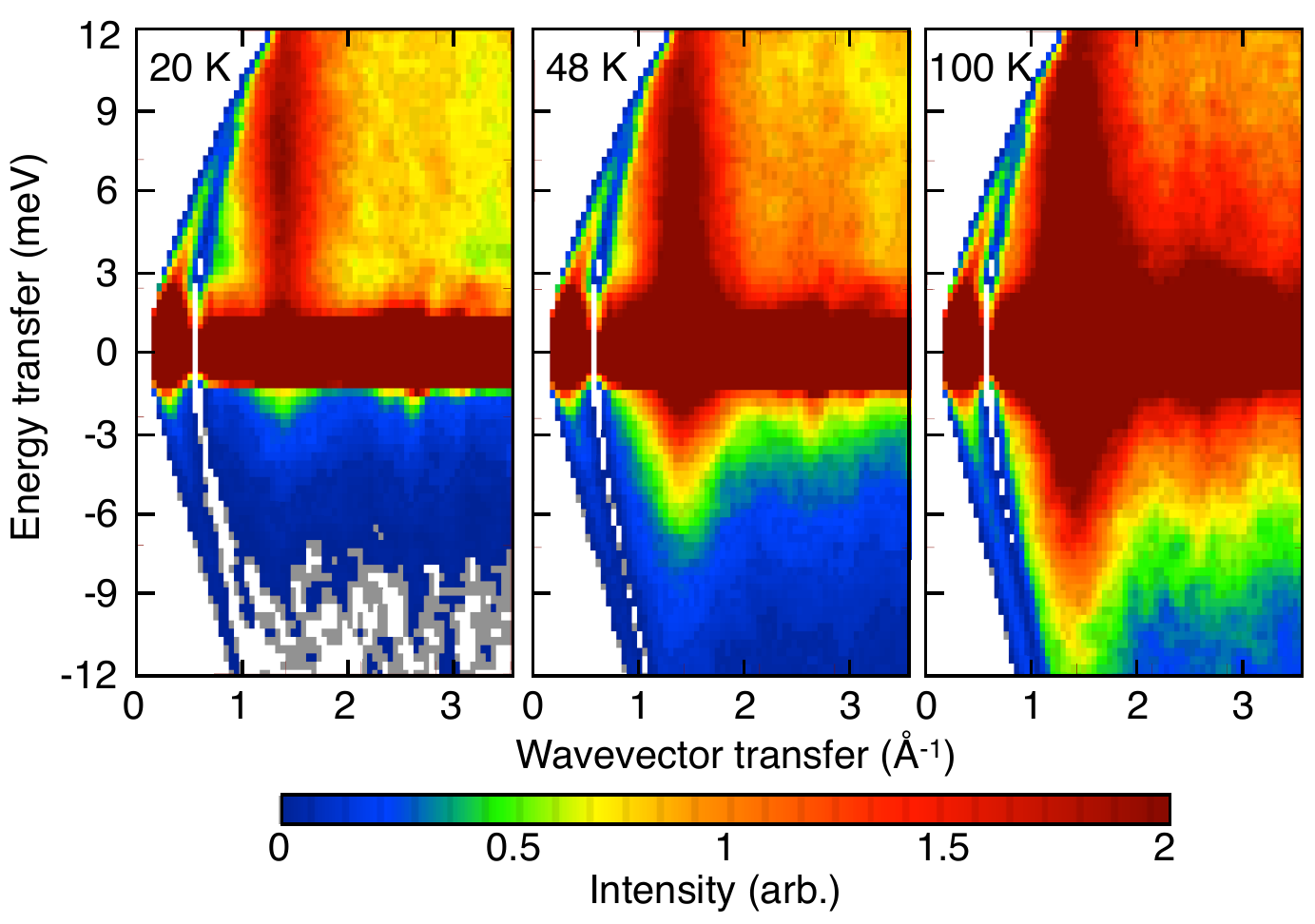}
	\caption{(Color online) Intensity maps of the scattering function $S(Q,\omega)$ as measured using  IN4 with an incident neutron energy of $E_{i}$=17~meV at three different temperatures, $T$=~20, 48 and 100~K, from left to right. The curved feature emanating from $Q\approx$ 0.7 \AA$^{-1}$ is due to the gap in the detector assembly at about $2\theta\approx$10\grad.}
	\label{sqw}
\end{figure}

The high energy dynamic response of the system as probed by thermal neutron TOF spectroscopy on the IN4 spectrometer shows many features not reported earlier.  An intense ridge of quasi-elastic magnetic scattering centered around $Q=1.4$~{\AA}$^{-1}$, at the same position as the peak in the ``static'' diffuse scattering measured in the more restrictive dynamic range of D7, is seen in the $(Q,\omega)$ plane as depicted in Fig.~\ref{sqw}.  This ridge extends over a broad energy range with a non--trivial temperature dependence which will be detailed below. The broad energy range over which the ridge is observed, and its diffuse nature, suggest that it is due to an ensemble of correlated magnetic ``clusters'' with a large size distribution and a macroscopic number of interacting degrees of freedom.

\begin{figure}
\includegraphics[width=3.3in]{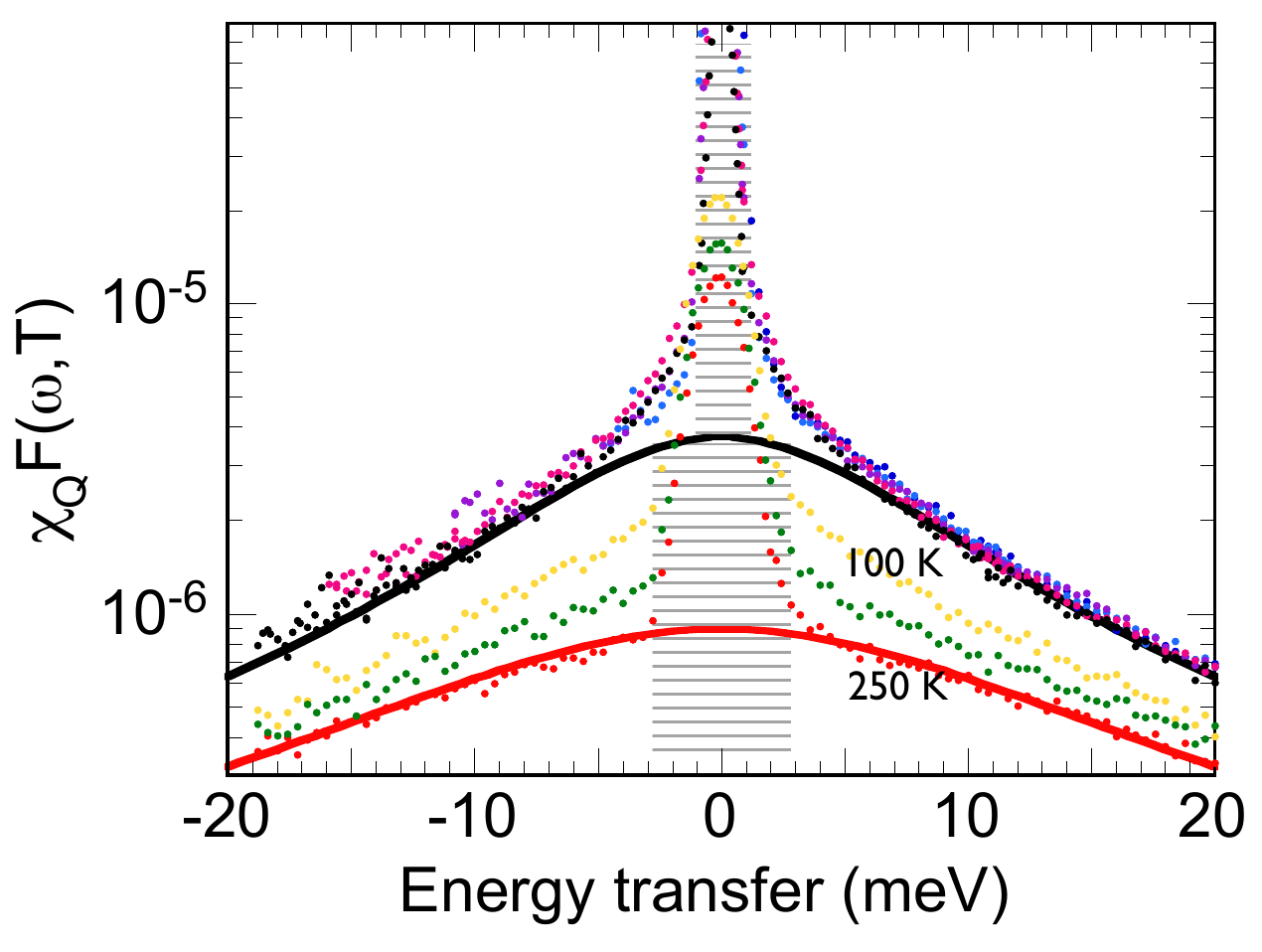}
\caption{(Color online) Temperature dependence of the experimentally evaluated spectral response at $Q=1.4$~\AA$^{-1}$ defined in Eq.~\ref{sqwt}. Data from bottom up taken at 250 K (red), 200 K (green), 150 K (yellow), 100 K (black), 70 K (pink), 48 K (purple), 20 K (light blue) and 1.5 K (dark blue).  The ranges of the data are limited on the energy gain side (negative energy transfer) by the thermal population/detailed balance factor and at T=1.5 K only the energy loss data is shown. Various instrument settings were used and combined.  The shaded area covers the instrumental resolution range $\pm$ 3 meV for $T>100$ K and $\pm$ 0.7 meV for $T\leq100$ K.  The solid lines show examples of fits to the data of a single broad lorentzian component at $T=250$ K (red solid line) and $T=100$ K (black solid line).  At $T=100$ K we note the onset of a narrow component that is examined in detail in Figs.~\ref{in5spectra} and~\ref{in5_results}, while the broad component stays virtually independent of temperature below 100 K.}
\label{fomegat}
\end{figure}

\begin{figure}
\includegraphics[width=3.3in]{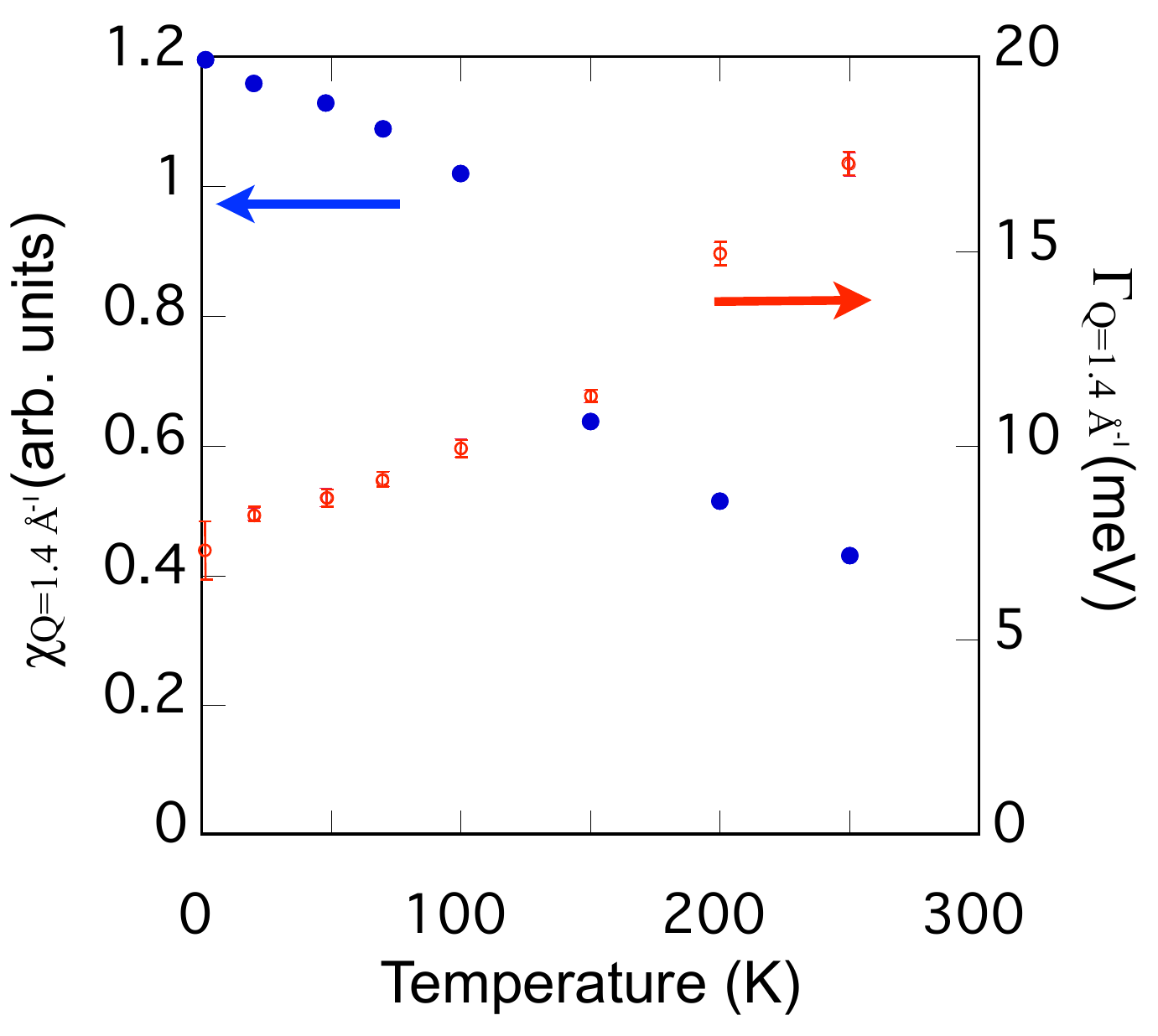}
\caption{The temperature dependence of the line--width $\Gamma$ (red hollow circles) and of the intensity $\chi$ (blue circles) at a wavevector transfer of $Q=1.4$ \AA$^{-1}$ (the peak of the susceptibility), of the broad lorentzian component of the inelastic response measured on IN4.}
\label{broadparams}
\end{figure}

We can examine the line--shape and temperature dependence of this magnetic scattering independently of any particular model by considering the following general expression for the scattering function, 
\begin{eqnarray}
\label{sqwt}
	S(Q,\omega,T) & = & (1-e^{-\frac{\hbar\omega}{kT}})^{-1}\omega\chi_{Q}(T)F_{Q}(\omega,T) \nonumber \\
	& = & (1-e^{-\frac{\hbar\omega}{kT}})^{-1}\chi''(Q,\omega,T)
\end{eqnarray}

Using this relationship  the product of the wave--vector transfer dependent susceptibility and the spectral weight function, $\chi_{Q}(T)F_{Q}(\omega,T)$, can be evaluated from the experimental data.  The spectral weight function is normalized,  $\int{F(\omega,T)d\omega=1}$, such that the temperature dependence of the integrated inelastic intensity is given by the $Q$--dependent susceptibility, $\chi_{Q}(T)$.  Fig.~\ref{fomegat} shows the temperature dependence of $\chi_{Q}(T)F_{Q}(\omega,T)$ taken from a constant $Q$--cut through $S(Q,\omega)$ centered at 1.4 \AA$^{-1}$, on cooling from $T$=280~K to $T$=1.5~K.  The continuous lines on the plot  follow a single lorentzian line--shape of the form 
\begin{equation}
	F_{Q}(\omega,T) =\frac{1}{\pi}\cdot\frac{1}{\omega^{2}+\Gamma_Q^{2}}
	\label{lorentz}
\end{equation} 
which describes the data at high temperatures rather accurately.  Furthermore we note that the observed narrowing of the line--shape as the temperature is reduced, appears to take place in the energy transfer range $|\hbar\omega|\lesssim{5}$~meV.  Therefore we proceed by restricting our attention to the high--frequency fluctuations, fitting only the broad part of the response outside of this range to obtain temperature dependent values of the broad lorentzian line--width, $\Gamma_Q$ and intensity $\chi_{Q}$.  These are depicted in Fig.~\ref{broadparams}. The line--width $\Gamma_Q$ at $Q=1.4$~\AA $^{-1}$ falls from $~\sim 20$~meV at 300 K to $\sim 10 $~meV at 100 K, and thereafter remains roughly constant with temperature, falling only slightly to $\sim 7$~meV at 1.5 K.  The peak susceptibility,  $\chi_{Q}$ at $Q=1.4$~\AA $^{-1}$ associated with this broad component, increases slowly between 300 K and 1.5 K.

\begin{figure}
\includegraphics[width=3.1in]{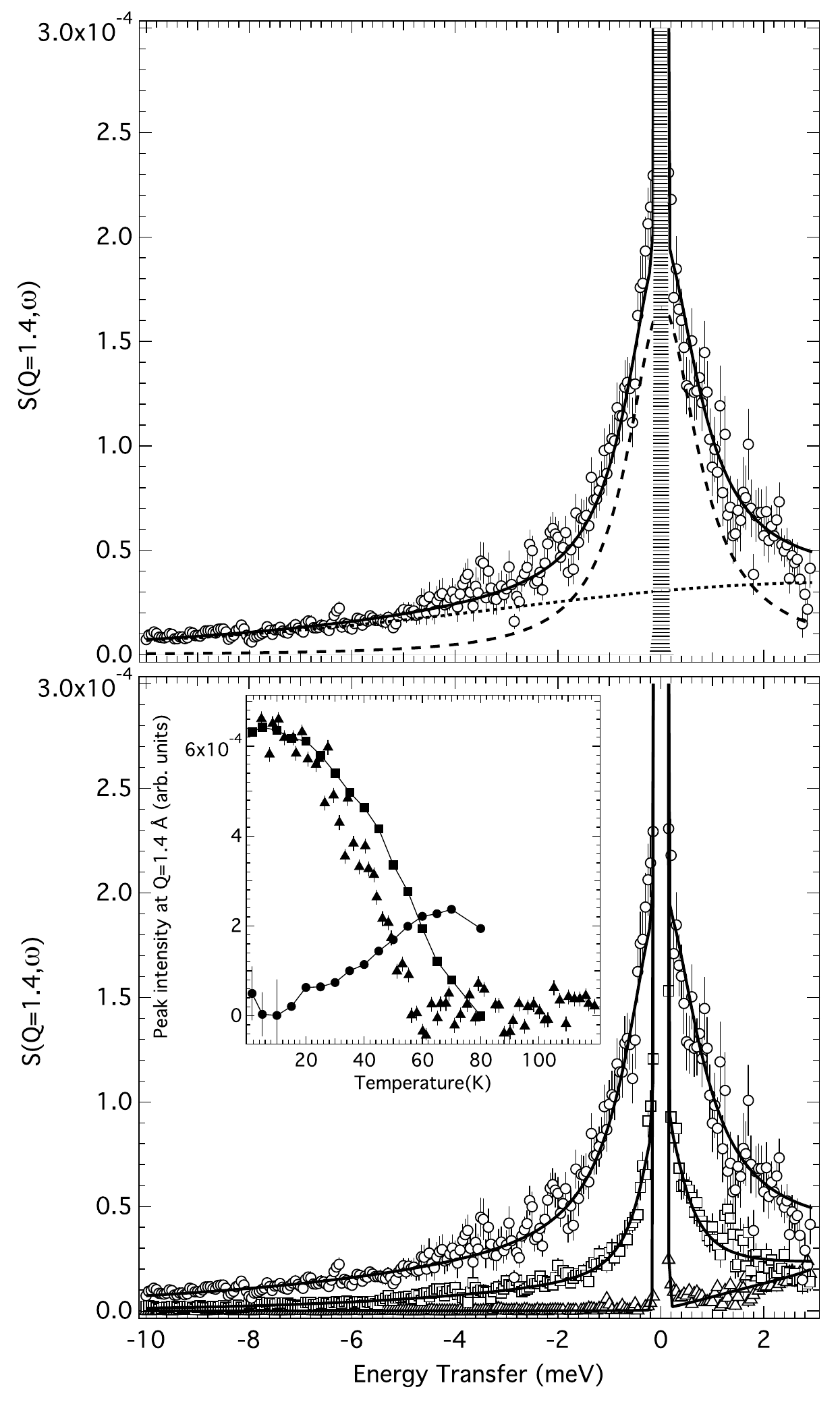}
\caption{(top panel) The inelastic response of \YCBCO\ at $Q=1.4$~\AA$^{-1}$ measured at 80~K on the IN5 spectrometer.   The solid line shows a fit to a two-lorentzian model as described in the main text.  The individual broad and narrow lorentzian components are shown as dotted and dashed lines respectively.  The instrumental resolution was modeled as a Gaussian of FWHM $\sim 120$ meV and is shown as the hatched area.
\\~~~(lower panel) Evolution of the inelastic response at $Q=1.4$~\AA$^{-1}$ as a function of temperature measured on IN5.  The temperatures shown are 80~K (circles), 40~K (squares) and 10~K (triangles).   The broad Lorentzian component is all that remains at low temperatures.  The inset shows the temperature evolution of the elastic (squares) and inelastic (circles) integrated intensity at $Q=1.4$~\AA$^{-1}$.  For comparison, the elastic intensity measured at much greater resolution (FWHM $\sim$ 1 $\mu$eV) on IN16 is also shown (triangles).  The scale is adjusted to highlight the relative difference with respect to data measured at the higher temperature range, in order to depict the development of the magnetic contribution.}
\label{in5spectra}
\end{figure}

\begin{figure}
	\includegraphics[width=3.3in]{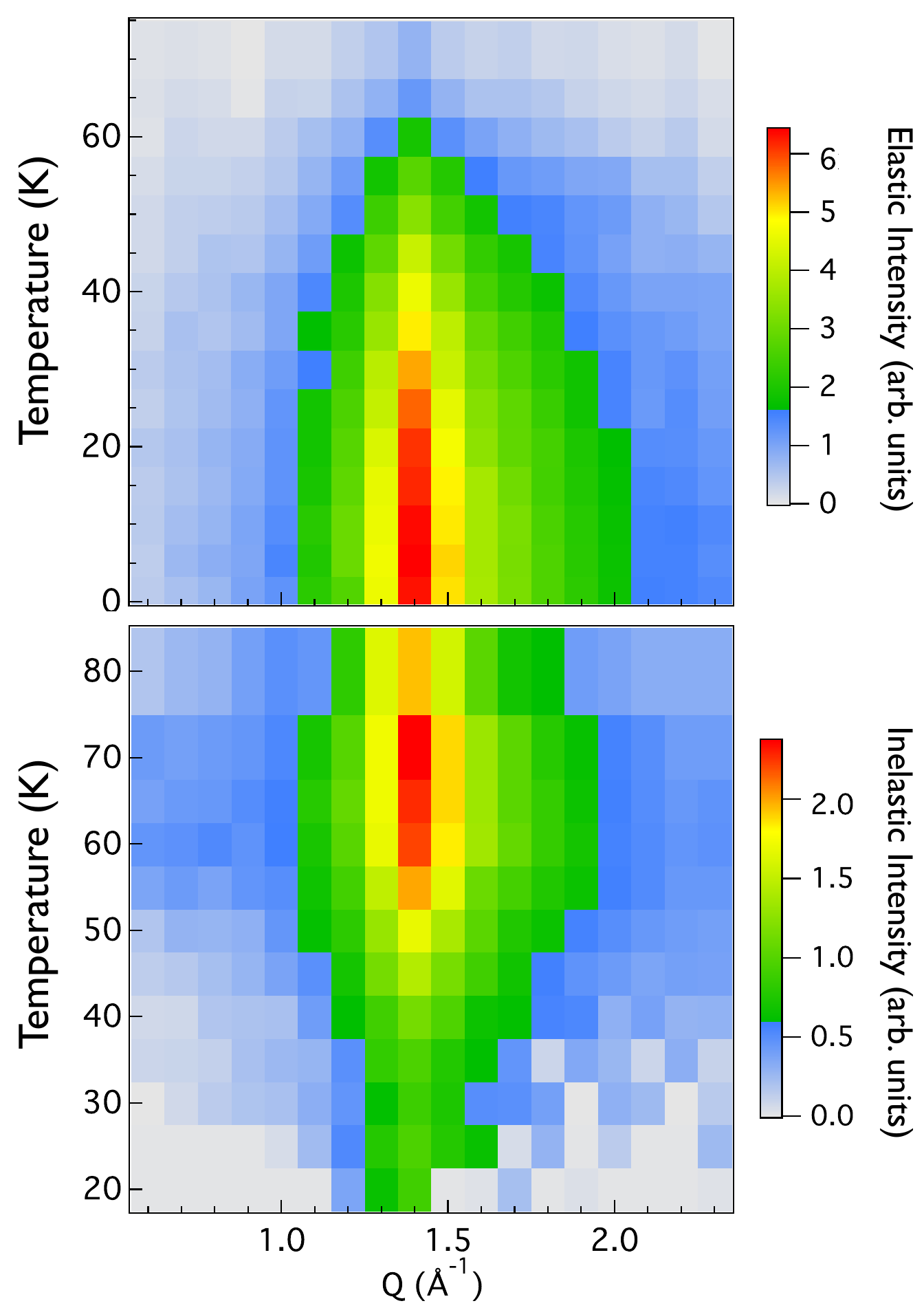}
	\caption{(Color online) The integrated ``elastic'' response (top panel) and integrated inelastic low--frequency ($\sim{1}$~meV) response (lower panel) of \YCBCO\ derived from fitting spectroscopic data from IN5 to Eq.~\ref{sqwt}, as a function of temperature and wavevector transfer.}
	\label{in5_results}
\end{figure}

In our judgement, the line--shape of the total response is complex and the existence of a distinct broad component is not justified by any self--evident arguments.  Nevertheless, considering that both $\Gamma_Q$ and $\chi_{Q}$ are only weakly temperature dependent for $T\lesssim{100}$~K, and that no great deviation in these parameters is observed around the spin freezing transition seen in our susceptibility data (Fig.~\ref{SuscGraph}), we proceed further to characterize the narrower central part of the response.  In order to accomplish this, we use data taken over a smaller dynamic range, but with better resolution using the IN5 spectrometer, shown in Fig.~\ref{in5spectra}.   Keeping the parameters of the broad component fixed ($\Gamma=10$~meV and $\chi_{Q}\sim{}1$) at temperatures below 80~K, we can fit the narrow response with another lorentzian form using Eqs.~\ref{sqwt} and \ref{lorentz}, shown in Fig.~\ref{in5spectra}.   The elastic intensity and the integrated low--frequency response, found from the fitted IN5 spectra as a function of $Q$ and $T$ are shown in Fig.~\ref{in5_results}.  ``Elastic'' refers to the part of the response entering within the energy resolution width of the instrument.  On IN5, this energy resolution was found to have a FWHM of $\Delta{E}\approx{120}$~$\mu$eV.  It can be seen from Fig.~\ref{in5_results} that the elastic and inelastic intensities follow roughly the same $Q$--dependence as the static magnetic structure factor presented in section~\ref{polar}.  The intensity of the low--frequency response peaks at $T\approx{70}$~K just above the macroscopic freezing temperature $T_{g}=48$~K (see inset of Fig.~\ref{in5spectra}).  The line--width remains small, of the order of $\Gamma'\sim{1}$~meV, independent of temperature.  At the lowest temperatures $T\le{20}$~K the low--frequency inelastic intensity of magnetic origin essentially disappears - an observation backed up by our neutron spin-echo data shown below. This temperature corresponds roughly to the activation temperature of 25--30 K derived from the specific heat measurements shown in Sec.~\ref{bulk}.  At the same time the elastic intensity increases smoothly below $T\approx{80}$~K, indicating formation of a static magnetic ground--state. This is a rather standard signature of a dynamical magnetic transition: a similar low--frequency response was also seen above the freezing of magnetic fluctuations in SrCr$_{9q}$Ga$_{12-9q}$O$_{19}$ (SCGO)~\cite{mond99a}. 

We gain further insight into this phenomenon using a so--called \emph{elastic scan} obtained on the backscattering spectrometer IN16.  We record the elastic scattering intensity $I_{\rm{el}}$ during a temperature sweep over a $Q$--range of 0.25~{\AA}$^{-1}\le{Q}\le{2}$~{\AA}$^{-1}$.  On IN16 the elastic energy resolution has a FWHM of $\Delta{E}\approx{1}$~$\mu$eV.  The elastic and low--frequency inelastic intensities at $Q=1.4$~{\AA}$^{-1}$ from both IN5 and IN16 are shown in the inset of the lower panel of Fig.~\ref{in5spectra}.  The overall trend in the behavior of the elastic response on IN5 and IN16 is very similar, indicating only a weak dependence on the characteristic time--scale of the probe~\cite{mur78a}, of the order of 10$^{9}$~Hz on IN16 and 10$^{11}$~Hz on IN5.  Similarly, comparison of the temperature dependences in the inset of Fig.~\ref{in5spectra} with the total integrated magnetic moment measured on D7 (which includes moments fluctuating at frequencies up to 10$^{12}$~Hz) shown in the inset of Fig.~\ref{RMC_Correlations}, shows that the spin--freezing behavior must be taking place over much longer timescales than those probed by the neutron measurements presented thus far.  Accordingly it is interesting to compare these results with neutron spin echo (NSE) spectroscopic measurements, which directly probe the time dependence over similar timescales.

\begin{figure}
\includegraphics[width=3.0in]{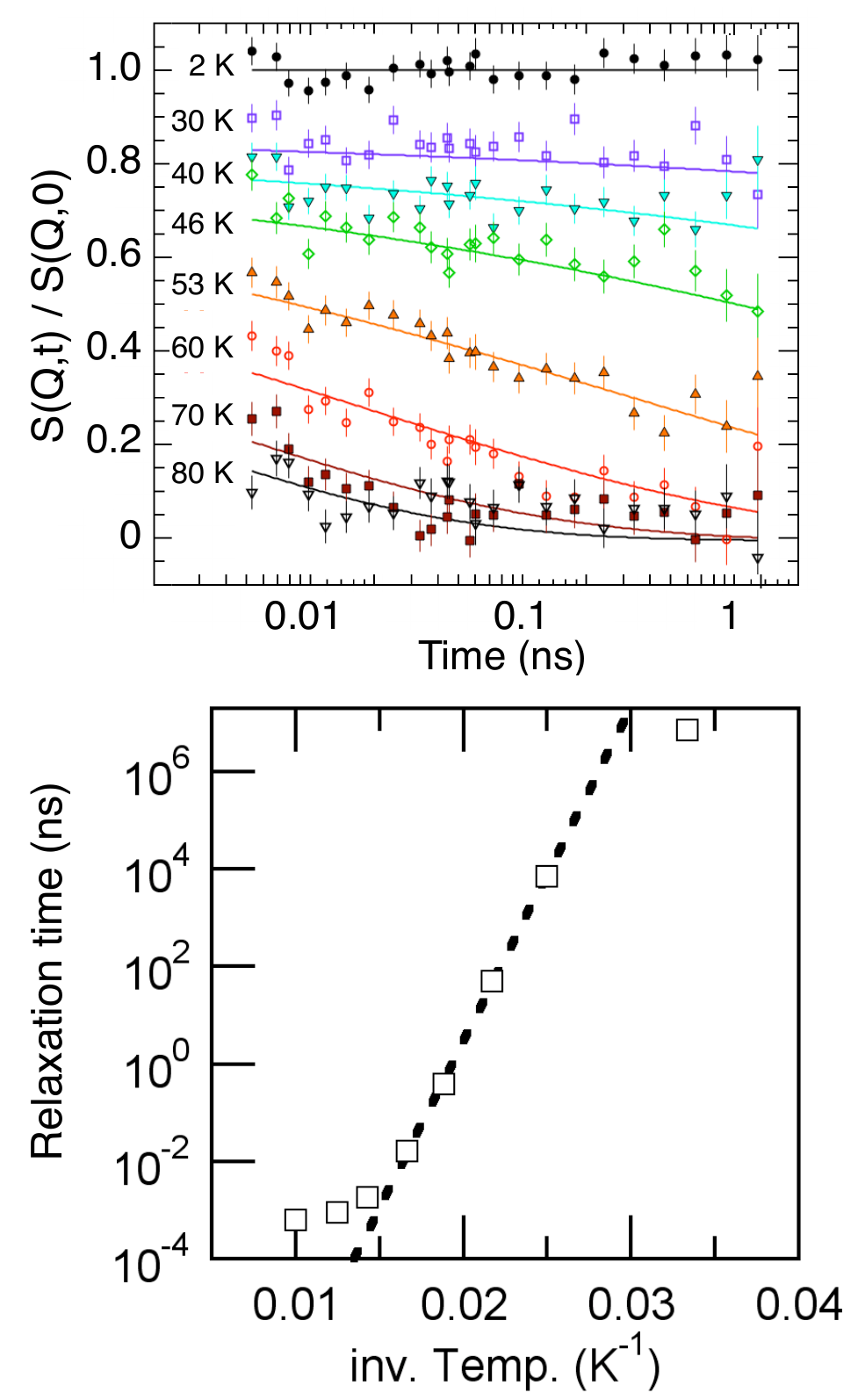}
\caption{(Color online) Upper panel - The intermediate scattering function on the correlation peak probed by NSE between 2~K and 80~K. Lower panel -  the fitted relaxation times (see text), an approximate Arrhenius behavior is indicated by the dotted line.}
\label{NSE_Result}
\end{figure}

The data obtained with NSE show a continuous slowing down when cooling across the macroscopic freezing temperature $T_{g}$.  This has been observed in various spin glasses and is not too surprising given the different time--scales probed by the different techniques ($f\sim{10}^{9}$~Hz for NSE, 0~Hz~$\le{f}\lesssim{10}^{3}$~Hz for DC or AC susceptibility).~\cite{ehlreview06,gardner01} At base temperature the response is truly elastic, $S(Q,t)/S(Q,0)=1$ over the full time window probed.  This is different with respect to the pyrochlore slab compounds SCGO and BSZCGO in which  a fast initial relaxation leads to $S(Q,t)/S(Q,0)\le{1}$, see Ref.~\onlinecite{mut06}. It also means that dynamics on the picosecond time--scale do not exist at all at base temperature which corroborates the result hinted at by the apparent disappearance of the low--frequency Lorentzian component at low temperatures as seen from the IN5 data, suggesting the absence of any magnetic fluctuations below $\hbar\omega\lesssim{1}$~meV, see Figs.~\ref{in5spectra} and~\ref{in5_results}.

On increasing the temperature, the NSE data show a response that is much more stretched out in time than a single exponential relaxation function and gradually shifts down through the dynamic window of the instrument.  This indicates that the response has contributions over many time scales simultaneously, making it difficult to fit a single time scale (relaxation time) to the data, considering that the dynamic range of the instrument is short by comparison, covering less than three orders of magnitude in time.  An attempt was made to quantify a characteristic time scale for the spin relaxation by fitting the NSE data to stretched exponential relaxation function, $S(Q,t)\propto\exp\left[-\left(t/\tau\right)^{\beta}\right]$.  The resulting relaxation times are shown in the lower panel of Fig.~\ref{NSE_Result} and appear to follow an Arrhenius law: $\tau=\tau_0\cdot\exp(\Delta/T)$ with an energy barrier of $\Delta\sim{1580}$~K (140~meV), without any diverging behavior at $T_{g}$.  The phenomenological $\beta$ parameter is very small, $\beta\sim{0.15}-0.30$, with a weak temperature dependence.


\section{Discussion}
\label{Discussion}

The results shown above give some new insights into the frustrated magnetic properties of the \YCBCO\ compound. The polarized neutron diffraction cross--section is in good quantitative agreement with the data reported earlier,~\cite{schweika07} although the conclusions we draw from our modeling of the data are somewhat different.  The most important new finding is the observation of an extended fast dynamic response that was missed up to now.  This allows us to abandon the earlier hypothesis of magnetism based on the kagome layers only, bringing our conclusions concerning the atomic structure and the magnetic short--range ordered structure into line with other published work on undoped \YBCO~\cite{chapon06,manuel09}. 

Our high--resolution neutron diffraction work reveals that the introduction of Ca into \YCBCO\ results in a disordered structure, with a splitting of the O1 and O3 sites, and a consequent distribution of magnetic exchange pathways, depending on whether the CoO$_4$ tetrahedra are connected to Ca or Y ions.  It is likely that this magnetic exchange disorder is responsible for the lack of long--range magnetic order in \YCBCO. The symmetry of the crystal is \emph{trigonal} and not hexagonal as was previously reported.~\cite{schweika07}  The introduction of atomic disorder and the fact that Ba$^{2+}$ is not under--bonded, may possibly be responsible for the lack of any signature of a trigonal-orthorhombic distortion in the magnetic susceptibility of \YCBCO\ -- although this has to be verified with low--temperature diffraction measurements.  

Sizable spin correlations develop below  $T\sim{100}$~K, leading to a well--developed short--range correlated but fluctuating state.  The magnetic response remains fully dynamic down to $T_{g}\approx50$~K. At this temperature a partial freezing sets in as evidenced by bulk susceptibility and high--resolution inelastic neutron scattering measurements.  Acknowledging the similar dynamic acceptance windows of D7 and IN11C (see section~\ref{Experimental}) we can conclude that the diffraction response obtained on D7 in the low--$T\lesssim{20}$~K range is essentially elastic scattering probing a static magnetic ground state.  Accordingly the RMC spin configuration applies only to the frozen state.  

We have analyzed the experimental static structure factor $S(Q)$ measured in the limited dynamic range of our cold neutron polarized diffraction measurement over the whole temperature range, from the low--$T$ partially frozen state up to $T=300$~K.  Using the RMC method, we have examined various hypotheses concerning the location and presence of magnetic moments.  Taking into account the deviation from ideal kagome geometry exhibited in the atomic positions of the $(6c)$--sites with exact and ``kagome averaged'' shell distances gives identical results, suggesting that the bond length differences in the non--perfect kagome plane are not an important factor in determining the ground--state structure.  The model restricting the magnetism to the kagome planes only results in a much poorer representation of the data than does the model with equal Co spins on \emph{all} the Co sites.  This indicates that there are no low--spin Co$^{3+}$ ions in \YCBCO\ as asserted by Schweika and co-workers.~\cite{schweika07}  Indeed it would be somewhat astonishing if low--spin Co ions existed within this crystal structure.  The low--spin configuration of Co$^{3+}$ is not allowed in tetrahedral symmetry since it is not possible to have the two $t_{2}$ electron spins antiparallel. A low--spin state would be stabilized if there were a strong site distortion which could perhaps lower one of the $t_{2}$ orbitals to such an extent that it would be possible to overcome the Hund's rule coupling.~\cite{holl09}  From our diffraction data there is no evidence for such a strong distortion.  Also, it was recently concluded from a combined experimental and theoretical x--ray absorption spectroscopy study of \YBCO\ that both the Co$^{3+}$ and Co$^{2+}$ are in the high--spin state in this material.~\cite{holl09}  Furthermore, the ideal stoichiometry of \YCBCO\ should be written (Y$^{3+}$)$_{0.5}$(Ca$^{2+}$)$_{0.5}$(Ba$^{2+}$)(Co$^{2+}$)$_{2.5}$(Co$^{3+}$)$_{1.5}$(O$^{2-}$)$_{7}$.  Note that the Co$^{2+}$ to Co$^{3+}$ ratio is 2.5/1.5=1.67.  In the $P31c$ crystal structure, 3/4 of all Co ions form the kagome layered Co2 $(6c)$--site with the remaining 1/4 located on the Co1 $(2a)$--sites between the kagome layers. Thus the Co2 to Co1 ratio (=3) is higher than the Co$^{2+}$ to Co$^{3+}$ ratio (in \YBCO\ both ratios are equal to 3).  This means that even if the Co1 $(2a)$--sites all have Co$^{3+}$, the kagome layers must  contain a distribution of Co$^{3+}$ and Co$^{2+}$ ions.  Therefore we can assert that \YCBCO\ is a mixed valence compound and the kagome layers must contain a distribution of mixed \textit{S}=3/2 and \textit{S}=2 spins.  This mixed valence character adds to the disorder already present due to the Ca substitution and consequent exchange pathway disorder.  This reasoning  highlights how the O-stoichiometry can influence the magnetic moment distribution on the Co sites and therefore play a critical role in determining the physical properties, in parallel to what was already concluded for YBaCo$_{4}$O$_{7+\delta}$ in Ref.~\onlinecite{maignan06}.  Oxygen stoichiometry and homogeneity may be the factors explaining some of the differences seen in the bulk data of our sample when compared with that reported earlier.~\cite{valldor06}

The polarized neutron diffraction data further reveal that the total moment obtained over the restricted experimental dynamic range is considerably reduced with respect to the the expected full moment, in agreement with our observations of a wide dynamic range of magnetic fluctuations using thermal neutron spectroscopy.  

The inelastic neutron measurements are consistent with fluctuating dynamics over a restricted spatial range with no signature of propagating fluctuations.  The presence of propagating modes has been predicted for a classical kagome spin system~\cite{robert08} -- but only at low temperatures coinciding with the frozen regime of  \YCBCO.  The dynamic response shows a complicated non--lorentzian line--shape. In the absence of a microscopic picture we have made an attempt to quantify the $T$--dependence by recognizing that at higher $T$ the response is rather well described by a single lorentzian form with a width approaching $\Gamma\sim{20}$~meV at $T=300$~K.  Below a temperature $T\approx{100}$~K the intensity of the broad response flattens off, with a reduced line--width of $\Gamma\lesssim{10}$~meV, however, at these temperatures a low--frequency response develops on top of the broad response.  We obtain a picture in which a slowing down of dynamics precedes the eventual spin freezing and finally leads to the appearance of a fully elastic response.  The appearance  of this elastic part is seen in the high--resolution time--of--flight, backscattering and neutron spin echo data.  The neutron spin echo results indicate a truly elastic response at base temperature, $T=2$~K.  We can conclude that at low temperatures the response of the spin system separates into two parts, one that freezes and another that remains dynamic.  With the data available, however, it is not possible to identify an underlying mechanism for this on the atomic/molecular level.  

It is worth mentioning here that the magnetic properties of \YCBCO\ bear some resemblance to those of the pyrochlore slab systems SCGO and BSZCGO~\cite{mond00,bonnet04,mut07} but with some qualitative and quantitative differences.  
The dynamical response of \YCBCO\ shows rather clearly definable temperature dependent  energy scales, in contrast to the $\omega/T$--scaling and power law bulk susceptibility seen in the pyrochlore slabs and more recently in herbertsmithite.~\cite{DeVries09,helt10}  It is also surprising that spin freezing in \YCBCO\ happens at much higher temperature in comparison with the overall energy scale.  The role of disorder involved in the mechanisms leading to glassy ground state in the frustrated systems has attracted recent attention~\cite{saund07,andre10}~and controlled studies on \YCBCO\ might bring some more insight on this issue.

\section{Conclusions}
In this paper we have reported results and analysis of bulk and neutron diffraction and spectroscopic measurements made on the \YCBCO\ compound -- proposed as a realization of classical kagome spin system.  We have examined the magnetic signal of this compound over a broad dynamic range from below a $\mu$eV up to a few tens of meV.  Our observation of a broad dynamic response explains why the total moment carried by the Co--ions was earlier assumed to be too small, leading to the conclusion that only the Co--ions residing on the kagome planes were magnetic.  We have proposed a revised interpretation for the magnetic diffraction data obtained with cold neutrons, pointing to a situation in which all Co atoms are magnetic.  In addition to the broad dynamic response with notable temperature dependence we have found conclusive evidence for a slowing down of part of the dynamic response, leading to a spin--freezing associated with the appearance of a truly elastic short--range correlated signal at the low temperature limit.  This spin--freezing is thought to come about due to the presence of exchange disorder between Co$^{2+}$ and Co$^{3+}$ ions in the vicinity of Y or Ca ions.  We conclude that \YCBCO\ is a highly frustrated antiferromagnet with an elastic response indicating a frozen component in the ground state.

\begin{acknowledgments}
The authors acknowledge helpful discussions with B. Canals, J. Robert, W. Schweika and in particular, L. Chapon. The authors are grateful for the local support staff at the ILL.  JRS thanks K. Knight for help with the neutron diffraction measurements.
CP thanks C. Reibel (Institut Charles Gerhardt, Montpellier, France) and J. Le Bideau for help with the specific heat measurements.  Research at Oak Ridge National Laboratory's Spallation Neutron Source was sponsored by the Scientific User Facilities Division, Office of Basic Energy Sciences, U. S. Department of Energy. 
\end{acknowledgments}

\bibliography{BIB-YCBCO}

\end{document}